\documentclass[aps,prb,twocolumn,a4paper]{revtex4}

\usepackage{graphicx,color}
\usepackage{ifthen}
\usepackage[pdftex,breaklinks=true,colorlinks=true]{hyperref}

\usepackage[utf8]{inputenc}

\usepackage{amssymb}
\usepackage{amsmath}
\usepackage{epsfig}
\usepackage{tikz}

\newcommand{\journ}[5]
{\ifthenelse{\equal{#1}{pr}}{
Phys. Rev. {\bf #2}, \href{http://link.aps.org/abstract/PR/v#2/e#3}{#3} (#4).}
{\ifthenelse{\equal{#1}{prl}}{
Phys. Rev. Lett. {\bf #2}, \href{http://link.aps.org/abstract/PRL/v#2/e#3}{#3} (#4).}
{\ifthenelse{\equal{#1}{prb}}{
Phys. Rev. B {\bf #2}, \href{http://link.aps.org/abstract/PRB/v#2/e#3}{#3} (#4).}
{\ifthenelse{\equal{#1}{prd}}{
Phys. Rev. B {\bf #2}, \href{http://link.aps.org/abstract/PRB/v#2/e#3}{#3} (#4).}
{\ifthenelse{\equal{#1}{pra}}{
Phys. Rev. A {\bf #2}, \href{http://link.aps.org/abstract/PRA/v#2/e#3}{#3} (#4).}
{\ifthenelse{\equal{#1}{arxiv}}{
preprint \href{http://arxiv.org/abs/#2.#3}{arXiv:#2.#3}.}
{\ifthenelse{\equal{#1}{rmp}}{
\rmp {\bf #2}, \href{http://link.aps.org/abstract/RMP/v#2/e#3}{#3} (#4).}
{\ifthenelse{\equal{#1}{cond-mat}}{
preprint \href{http://arxiv.org/abs/cond-mat/#2}{cond-mat/#2}.}
{\ifthenelse{\equal{#1}{pre}}{
Phys. Rev. E {\bf #2}, \href{http://link.aps.org/abstract/PRE/v#2/e#3}{#3} (#4).}
{#1 {\bf #2}, #3 (#4).}}}}}}}}
}}

\newcommand{\journdoi}[6]{#1\ {\bf #2}, \href{http://dx.doi.org/#5}{#3} (#4).}

\newcommand{\la}{\langle}
\newcommand{\ra}{\rangle}
\begin{document}

\title{Non-equilibrium steady states in the quantum XXZ spin chain}

\author{Thiago Sabetta and Gr\'egoire Misguich}

\affiliation{Institut de Physique Th\'eorique,
CEA, IPhT, CNRS, URA 2306, F-91191 Gif-sur-Yvette, France.}

\begin{abstract}

We investigate the real-time dynamics of a critical spin-1/2 chain (XXZ model) prepared in an inhomogeneous initial state with different
magnetizations on
the left and right halves. Using the time-evolving block decimation (TEBD) method, 
we follow the front propagation by measuring the magnetization and entanglement entropy profiles.
At long times, as in the free fermion case (Antal {\it et al.} 1999), a large central region develops where correlations become time-independent and translation invariant.
The shape and speed  of the fronts is studied numerically and we evaluate the stationary current as a function of initial magnetic field and as a function of the anisotropy $\Delta$.
We compare the results with the conductance of a Tomonaga-Luttinger liquid, and with the exact free-fermion solution at $\Delta=0$.
We also investigate the two-point correlations in the stationary region and find a good agreement with the ``twisted'' form obtained
by J.~Lancaster and A.~Mitra (2010) using bosonization. Some deviations are nevertheless observed for strong currents.

\end{abstract}

\date{\today}

\maketitle

\section{Introduction}
Quantum quenches has become an active field of research,\cite{polko11} in part due to their experimental feasibility in cold atoms systems.\cite{bdz08}
These quenches also offer an interesting framework to address basic questions about non-equilibrium phenomena in general and transport or thermalization in isolated systems in particular.
Quantum quenches in spin chains have also been used to benchmark powerful real-time simulation methods
such as adaptive time-dependent density matrix renormalization group (DMRG) \cite{gkss05} or Time-evolving block decimation (TEBD).\cite{vidal04}

In this paper we focus on the steady states which emerge when a gapless quantum spin chain
\begin{equation}
 H=-\sum_i \left(S^x_i S^x_{i+1} + S^y_i S^y_{i+1}\right)-\Delta\sum_i S^z_i S^z_{i+1}
\label{eq:HXXZ}
\end{equation}
is initially prepared in a ``domain-wall'' state with different magnetizations, say $\langle S^z_i\rangle=\pm m_0$, on the left and on the right halves of the chain.
For instance, we may chose the initial state  to be the ground state of the following
Hamiltonian:\footnote{As we shall see in Sec.~\ref{ssec:step}, a spatially smooth external field (use of $\tanh$ function) does not
modify the stationary state properties compared to a sharp step function, but it simplifies the numerical simulations by reducing the amount of quantum
entanglement between the left and right fronts.}
\begin{eqnarray}
H_{t<0}&=&H+h \sum_{i=-L/2}^{L/2-1} S^z_i \tanh(i)
\label{eq:tanh}
\end{eqnarray}
and the evolution for $t>0$ is computed using $H$.
The study of such type of initial conditions has been initiated by Antal {\it et al.} \cite{antal99} (free fermion model) and have been extended in
many directions since then:
numerics on the XXZ chain,\cite{gkss05,sm11,ecj12}
dynamics of the formation of a quasi-long range correlations,\cite{rm04}
initial state with magnetic field gradient,\cite{lm10}
bosonization or continuum limit,\cite{lm10,lgm10}
Bethe Ansatz approach (gapped case),\cite{mc10}
influence of an additional defect at the origin,\cite{barthel13}
or full counting statistics in the free fermion case.\cite{er13}
The case where the initial state has different temperatures on both sides was also considered.\cite{ogata02,kp09,grw10,bd12}
We note that from of the point of view of the energy, this quench is a global one since
the initial state has a finite {\it energy density} above the ground state of Eq.~\ref{eq:HXXZ}. On the other hand, it may be considered as
local since, far from the origin ($i=0$), the initial state matches a magnetized {\it eigenstate} of Eq.~\ref{eq:HXXZ}.

When the chain is gapless ($|\Delta|\leq1$)  a central region  of length $l$ expands ballistically with time $t$. Inside this region, 
the correlations become independent of time.
In the limit of long times (but with $l$ still small compared to the total length $L$ of the chain) we thus have a large segment
where a steady state can be observed. These correlations are  not that of the ground state since the central region support some particle current flowing from the left to the right.
One aim of this work is to characterize this non-equilibrium steady-state (NESS)
and to make contact with some transport properties of the chain, such as the conductance. 
This approaches simply follows the unitary evolution of a finite and isolated system,
and does not make use of Lagrange multipliers to force some particle and/or energy currents through the system\cite{antal97,antal98}
nor make use of a Lindblad equation to describe the couplings to reservoirs.\cite{prosen2011,pss12,kps13}

The paper is organized as follows. In Sec.~\ref{sec:ff} we review some results about the noninteracting case ($\Delta=0$).
Sec.~\ref{sec:boso}  is a summary of the results of a bosonization approach in the interacting case.
In Sec.~\ref{sec:tebd} we present the numerical result for the front propagation, and
Sec.~\ref{sec:EE} deals with the dynamics of the entanglement entropy. In Sec.~\ref{sec:current}
we analyze the evolution of the magnetization current and the discuss how it is related transport (conductance).
Sec.~\ref{sec:correl} is an analysis of the correlation functions in the NESS. Secs.~\ref{sec:current}
and \ref{sec:correl} contain some discussion of the validity and limitations of
the bosonization approach for this problem. The last section provides a summary and discusses some future directions.
A derivation of the noninteracting NESS for a general initial state is given in Appendix ~\ref{sec:free} and \ref{a:residue}.

\section{Free fermion case $\Delta=0$}
\label{sec:ff}

The dynamics and the steady state
are  relatively well understood\cite{antal99} in the case of the $XX$ chain ($\Delta=0$), which reduces to a free fermion problem.
There, the growing central region with zero magnetization is bounded by two fronts propagating to the left and to the right.
The width of each front also grows with time since the leading edge of the front has a speed $v_{\rm max}$ which is larger than that of the ``back'' of the front, which propagates
at a lower speed $v_{\rm min}$. These two speeds are $v_{\rm max}=1$ and $v_{\rm min}=cos(\pi m_0)$.
It is only in  the limit of an infinitesimally small bias $|m_0|\ll1$
that $v_{\rm min}$ and $v_{\rm max}$ coincide with the Fermi velocity.\footnote{This is easy to understand since the initial state is then a low-energy perturbation of the half-filled Fermi sea, and therefore only contain particle-hole excitations close to the Fermi level.
On the other hand, a strong bias must involve fermionic excitations in a larger portion of the Brillouin zone and will contain
particles and holes with slower group velocities. These slower particles are responsible for the ``back'' of the front.}
Unless we start
from a saturated initial state ({\it i.e.} $m_0=\frac{1}{2}$), $v_{\rm min}$ remains non zero and the central region extends ballistically.
The scaling form of the front has been derived for free fermions\cite{antal99} and it is interesting to note that a simple hydrodynamic/kinematic argument\cite{antal08} is able also give the exact shape of the front,
including the exact values of the velocity $v_{\rm min}$ of the back of the fronts (this argument will be reviewed in Sec.~\ref{ssec:long_times}).

At long times ($1\ll v_{\rm min}t$ and $v_{\rm max} t\ll L$) the NESS occupies a large region of the chain 
and becomes asymptotically  translation invariant.
It can thus be described by its occupation numbers in Fourier space, $\langle c^\dag_p c_p\rangle_{\rm NESS}$.
In the case of Antal's quench, the result is:
\begin{equation}
 \langle c^\dag_p c_p\rangle_{\rm NESS}=
 \left\{\begin{array}{cl}
1&{\rm for}\;    p\in[-k_F^-,k_F^+] \\
0&{\rm otherwise}
\end{array}\right.
\label{eq:boosted_FS}
\end{equation}
with
\begin{equation}
 k_F^\pm=\pi(\frac{1}{2}\pm m_0).\label{eq:kfpm}
\end{equation}

We note that $k_F^+$ is the Fermi momentum of the left reservoir at $t=0$, while $k_F^-$ is that of the right one.
For a general initial state at $t=0$ one can show
in the noninteracting case that $\langle c^\dag_p c_p\rangle_{\rm NESS}$ is equal to
the occupation number $\langle c^\dag_p c_p\rangle_{\rm L}$ of the left reservoir if $p>0$
and is 
is equal to $\langle c^\dag_p c_p\rangle_{\rm R}$ (that of
the right reservoir) if $p<0$ (Fig.~\ref{fig:scheme-occupation}). This complete decoupling between left movers and right movers has already been obtained by explicit calculations for a few specific initial states,\cite{antal99,grw10}
but we give a simple and general microscopic derivation of this result in Appendix \ref{sec:free}. 
This results also agrees with the hydrodynamic description developed in Ref.~\onlinecite{antal97}. This steady state is a simple example of an athermal state
(which nevertheless has the form of a generalized Gibbs state,\cite{rigol} see Sec.~\ref{ssec:gge}).

\section{Tomonaga-Luttinger liquid physics and bosonization}
\label{sec:boso}

The steady state which develops in presence of interactions ($\Delta\ne 0$) is not known exactly, even though some
rigorous results have been obtained concerning some other NESS in the XXZ chain.\cite{prosen2011,kps13}
We begin by a brief summary of the results obtained in a bosonization approach.

Lancaster and Mitra\cite{lm10} have  used bosonization to study the spin dynamics from a  initial domain-wall state.
This continuum limit retains a single velocity in the problem (linearized dispersion relation),
and the detailed shape of the front is therefore not  captured. In particular, the fact that the fronts widen ($v_{\rm min}<v_{\rm max}$)
with time is not reproduced.
One important result is however the simple form of the correlations in the NESS region:
\begin{eqnarray}
    \langle S^+_{x+n} S^-_{x}\rangle_{\rm NESS} &=& \langle S^+_{x+n} S^-_{x}\rangle_{\rm gs} e^{- i \theta n}
 \label{eq:SpSmtwist} \\
    \langle S^z_{x+n} S^z_{x}\rangle_{\rm NESS} &=&\langle S^z_{x+n} S^z_{x}\rangle_{\rm gs} .
\label{eq:SzSztwist}
\end{eqnarray}
Here the initial state is the ground state of Eq.~\ref{eq:tanh}, 	with an external magnetic field such that the magnetization
is $\pm m_0$ far from the origin ($x\to \mp \infty$).
$\langle \cdots \rangle_{\rm gs}$ denotes the expectation values in the ground state of the homogeneous chain at zero magnetization.
The (nearest-neighbor) angle $\theta$ which describes the ``twist'' between the steady-state and ground state correlations
is given by:
\begin{equation}
	\theta= \frac{\pi m_0}{K} \label{eq:theta_K}
\end{equation}
where $K$ is the Luttinger parameter. For the XXZ chain (in zero external magnetic field), $K$  is a known function of the anisotropy
$\Delta$:\cite{lp75,affleck88}
\begin{equation}
	K^{-1}=\frac{2}{\pi}\arccos{\Delta}
	\label{eq:KDelta}
\end{equation} 
This bosonization result  for correlation functions (Eqs.~\ref{eq:SpSmtwist},\ref{eq:SzSztwist} and \ref{eq:theta_K}) is remarkable,
since the correlations appear to be almost identical to that of the ground state,\footnote{Such simple relation to the ground state correlations does of course not always hold: if the interaction strength $\Delta$ changes during the quench, a TL
liquid may acquire algebraically decaying correlations with exponents which are {\it not} the equilibrium ones.\cite{ic09,lm10}}
a somewhat uncommon situation in the context of quenches.

Although qualified as ``intriguing''\cite{lm10} and often interpreted as a spatial inhomogeneity and an absence of equilibration,
the oscillatory factor is a particularly simple way to introduce some particle (spin) current on the top of the ground state correlations.
Such oscillations are  already present in the free fermion case\cite{antal98,antal99}
since  Eq.~\ref{eq:boosted_FS} is equivalent the usual  (non-shifted) half-filled Fermi sea, but with modified fermions operators:
$\tilde c^\dag_x = c^\dag_x e^{i x \pi m_0}$. In momentum space this redefinition of the fermion operators
amounts to a simple shift: $\tilde c^\dag_p=c^\dag_{p+m_0/\pi}$.
This observation immediately leads to Eqs.~\ref{eq:SpSmtwist},\ref{eq:SzSztwist} with 
with $\theta=m_0/\pi$ in the noninteracting case ($K=1$ at $\Delta=0$).
From this point of view, the phase factor $e^{- i \theta n}$ is a the direct consequence of having more spinons going to the right than going to the left. 
Since bosonization is a long-wavelength and low-energy approach, we may expect this form  to hold at {\it long distances},
at least for small initial bias $m_0$ where the current carrying state is a low-energy state.
We will test this result numerically in the next section. Somewhat surprisingly, we will find in Sec.~\ref{sec:correl} that Eq.~\ref{eq:SpSmtwist} holds quite accurately
even at {\it short distances} and for finite bias.

\section{Front propagation}
\label{sec:tebd}

\subsection{Velocities}
\label{ssec:velo}
\begin{figure}
  \includegraphics[width=6cm]{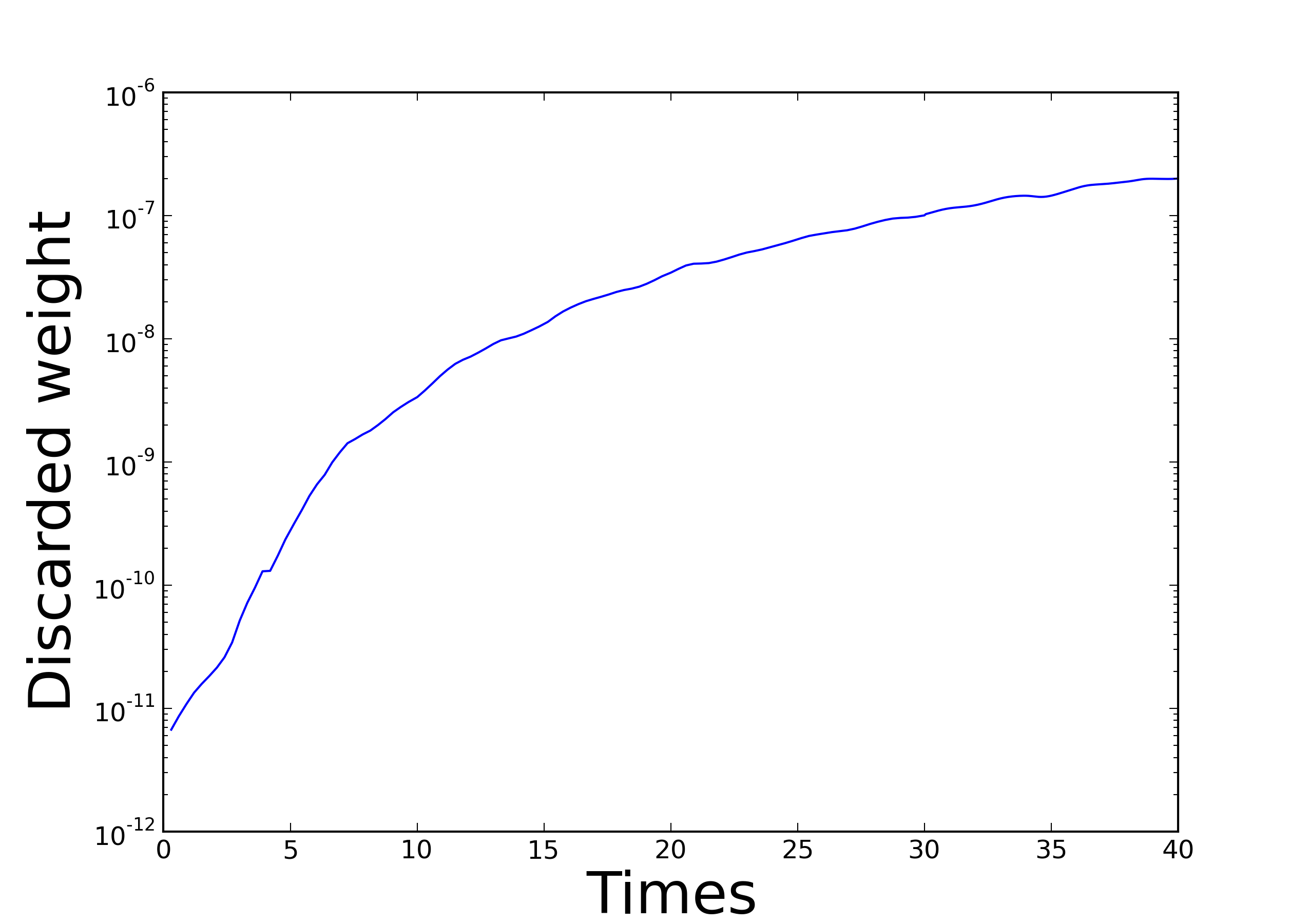}
  \caption[99]{(Color online) Weight of the discarded Schmidt eigenvalues during the TEBD truncation at each time step.
  The weight is summed over every link of the chain for each time step
  The figure shows the data for an anisotropy parameter $\Delta = 0.4$ and an initial magnetic field of $75\%$ of the saturation value ($h = f_{\rm sat}(1- \Delta)$, $f_{\rm sat}=0.75$).
}
\label{fig:discardedweight}
\end{figure}

We use the  (TEBD) to compute the time evolution from a domain-wall state.
Our code is based on the Open TEBD software.\cite{opentebd}
Unless specified otherwise, the chain has length $L=80$ sites and the wave-function are encoded using matrices of size $\chi=100$.
The initial state is 
chosen to be the ground state of the XXZ Hamiltonian with open boundary conditions and a  spatially varying magnetic field in the $z$ direction
(Eq.~\ref{eq:tanh}). 
As discussed in Sec.~\ref{sec:EE}, the amount of entanglement generated by these quenches is rather low and we checked that this value of $\chi$ is
sufficient to insure a good precision on all the observables discussed here.
As an illustration, Fig.~\ref{fig:discardedweight} shows the discarded weight
measured at each time step and at each truncation of the spectrum of the reduced density matrices. It remains relatively small during all the time evolution of the system (a few $10^{-7 }$ at most).

\begin{figure}
\includegraphics[width=9cm]{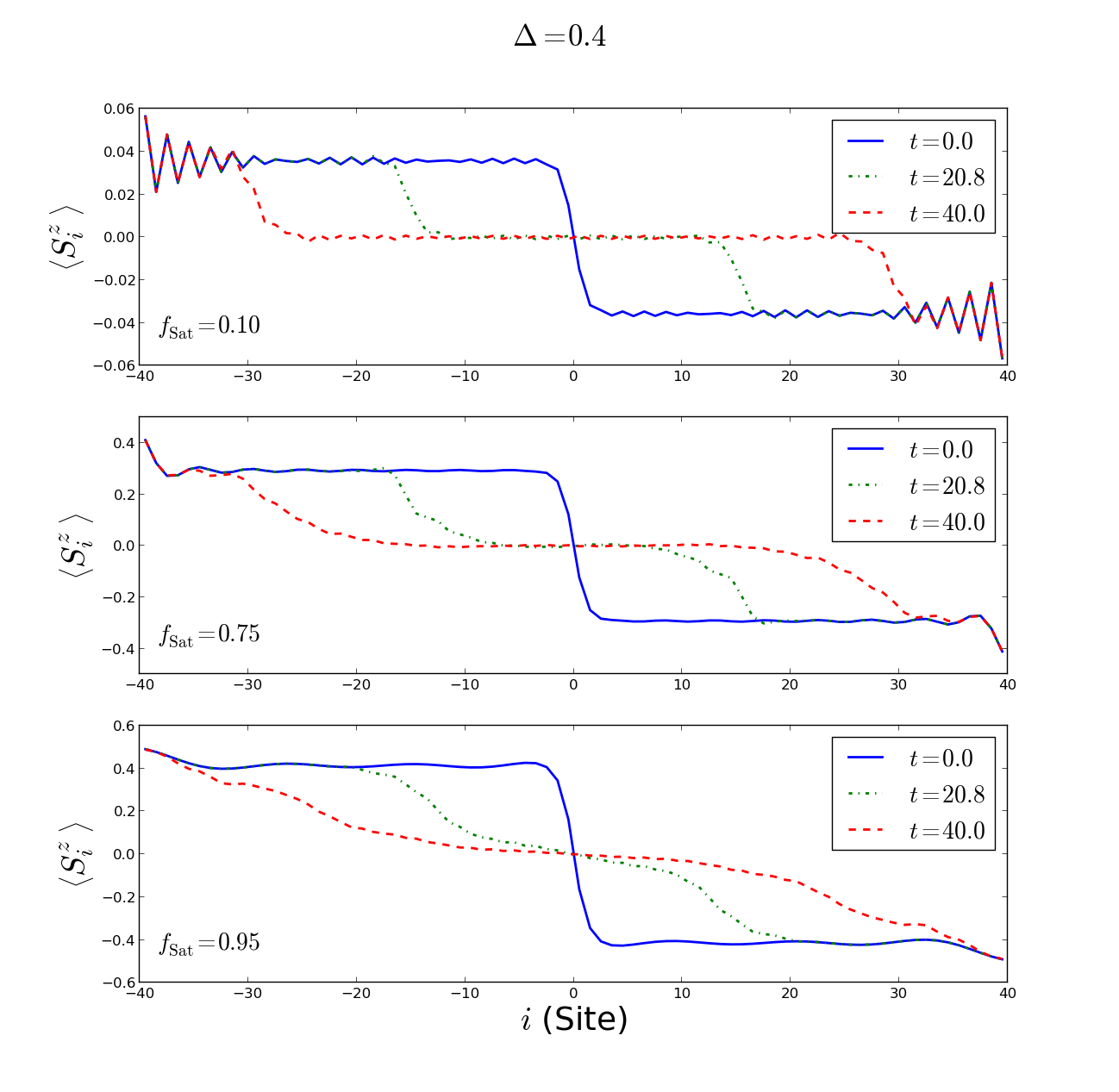}
\caption[99]{(Color online) Magnetization as a function of site position $i$ at the initial time (blue) and at $t=40$ (red).
Anisotropy parameter $\Delta=0.4$. The different panels correspond to different amplitudes of the initial magnetic field $h$, and hence different 
height of the initial magnetization steps $h=f_{\rm sat}h_{\rm sat}(\Delta)$, where $h_{\rm sat}=1-\Delta$.
From top to bottom $f_{\rm sat}=0.1,0.75$ and $0.95$. 
}
\label{fig:Sz}
\end{figure}

Typical  initial magnetization profiles are shown in Fig.~\ref{fig:Sz}.
At $t=0$ (blue lines) the external magnetic field is switched off and the time evolution is performed according to $H$ only.
Note that contrary to Ref.~\onlinecite{lm10}, the value of $\Delta$ is not changed between $t<0$ and $t>0$.

\begin{figure}
\hspace*{-0.3cm}\includegraphics[width=13.8cm]{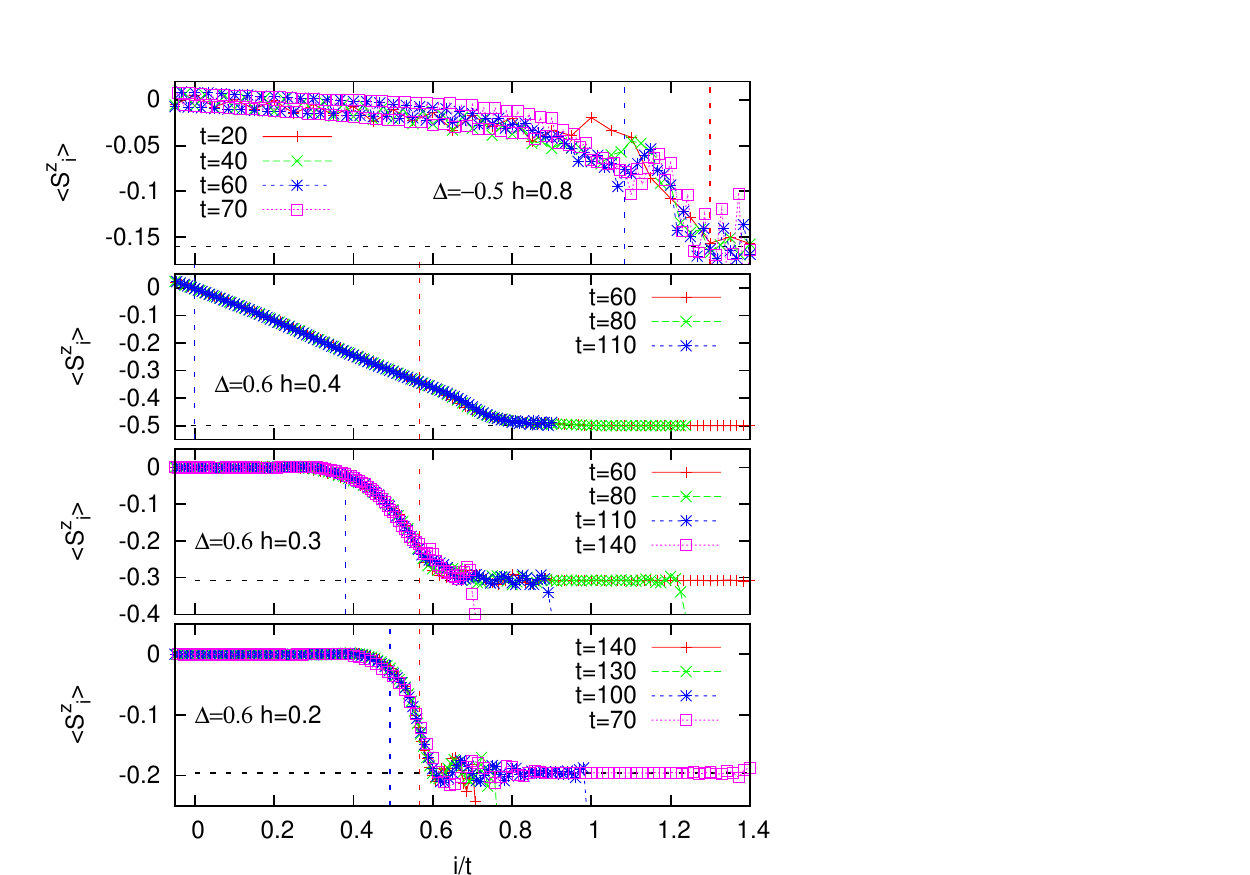}
\caption[99]{(Color online) Magnetization front in a chain of length $L=200$ sites (keeping respectively 150, 100,250 and 150 states from top to bottom).
Top: $\Delta=-0.5$, $h=0.8$ ($m_0\simeq 0.16$, indicated by an horizontal dashed line).
2nd: $\Delta=0.6$, $h=0.4$ ($m_0=0.5$).
3rd: $\Delta=0.6$, $h=0.3$ ($m_0=0.30$).
Bottom: $\Delta=0.6$, $h=0.2$  ($m_0\simeq 0.196$).
The position $i$ is scaled with time $t$
to make apparent the emergence of a limiting front shape.
A dashed red vertical line indicates the velocity of elementary excitations in an homogeneous XXZ chain at $\langle S^z \rangle=0$ (given by Eq.~\ref{eq:vmax}) and
a dashed blue line indicates that velocity for $\langle S^z \rangle=m_0$
(from the numerical solution of the Bethe Ansatz equations in presence of an external magnetic field\cite{BetheEquations}).
}
\label{fig:profile200}
\end{figure}

As already noted in Ref.~\onlinecite{gkss05} (where $m_0=\frac{1}{2}$), the domain-wall  quench in an XXZ model  gives rise to a ballistic front propagation.
As for the free-fermion case, the magnetization profile $\langle S^z_r\rangle$ acquires a limiting shape when the position $r$ is rescaled by time.
Fig.~\ref{fig:profile200} indeed shows a reasonably good collapse of the magnetization curves computed at different times.

The front region is characterized by two different velocities: the
leading edge of the front propagates at a velocity $v_{\rm max}$ which is larger than the velocity $v_{\rm min}$ of the back of the front.
In the free fermion case these velocities correspond respectively to the Fermi velocities at magnetization $0$, and $m_0$.
In presence of interactions ($\Delta\ne0$) the analog of the Fermi velocity is the group velocity of the spin excitations.
At zero magnetization this velocity is a known function of $\Delta$:\cite{dg66}
\begin{eqnarray}
 v&=&\frac{\pi}{2}\sin(\gamma)/\gamma \label{eq:vmax} \\
 \cos(\gamma)&=&-\Delta.
\end{eqnarray}
The above velocity corresponds to the red vertical lines in Fig.~\ref{fig:profile200} (see also the inset of Fig.~\ref{fig:profile400}).
We observe a reasonable agreement with the location of the leading edge of the front.
Since finite-size effects are presumably still important, the front velocity $v_{\rm max}$  may exactly coincide with
the spinon velocity of Eq.~\ref{eq:vmax} in the thermodynamic and long-time limit. 
A front propagation at this velocity has also been observed in a different quench of the XXZ spin chain.\cite{lhmh11}
Still, we note that on the present data the front seems
to propagate slightly faster than Eq.~\ref{eq:vmax} (case $m_0=\frac{1}{2}$ in particular).

At finite magnetization there is no closed formula for the group velocity of the spin excitations of the XXZ chain,
but it can be determined by solving numerically some integral equations.\cite{BetheEquations} The result corresponds
to the vertical blue lines in Fig.~\ref{fig:profile200}. Again the agreement with the front locations is reasonable but
it is not clear whether this velocity of the excitations in the homogeneous chain matches that $v_{\rm min}$ of the back of the front.
In the fully polarized case ($f_{\rm sat}=1$) we get an almost perfect collapse of the different front profiles. In this particular case the back of the front stays at the origin ($r/t=0$)
indicating clearly that the lower velocity $v_{\rm min}$ vanishes in this case where $m_0=0$ (as for free fermions).

\subsection{Density oscillations}
\label{ssec:osci}
\begin{figure}
\includegraphics[width=9cm]{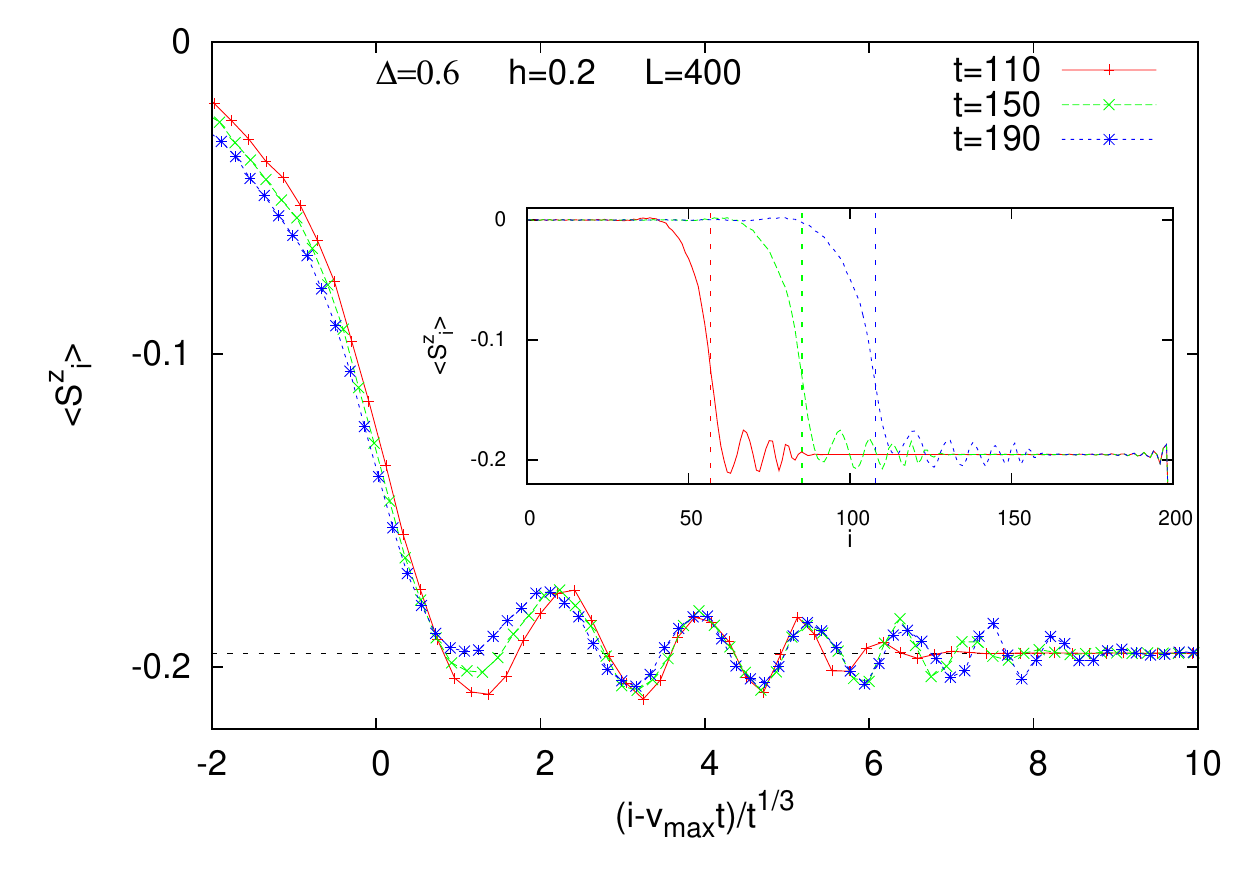}
\caption[99]{(Color online) Scaling of the oscillations in the magnetization front in a chain of length $L=400$ sites (keeping 200 states)
with $\Delta=0.6$, $h=0.2$  ($m_0\simeq 0.196$).
In the main panel the position $i$ has been shifted by $tv_{\rm max}$, where $v_{\rm max}\simeq 0.56751$ is an estimate of the front velocity from
the spinon group velocity in an uniform chain at zero magnetization and interaction strength $\Delta=0.6$.
The distance to the edge of the front has then been rescaled by a factor $t^{-1/3}$
to make contact with the oscillations observed in the noninteracting case.\cite{hf04}
In the inset the magnetization $\langle S^z_i\rangle$ is shown without any rescaling of the position $i$.
The red, green and blue dashed vertical lines correspond to estimates of the front-edge locations $i=t v_{\rm max}$
at times $t=100$, 150 and 190 respectively.
}
\label{fig:profile400}
\end{figure}

To conclude this section devoted to the magnetization fronts, we describe the density oscillations present ahead of the front.
These oscillations are visible in Fig.~\ref{fig:profile200}, and are magnified in Fig.~\ref{fig:profile400} for
a longer chain (400 sites) and longer times (up to $t=190$). The spatial period of these oscillations grows with time, but probably more slowly than $t$.
This should be compared with the $t^{1/3}$ law that is present in the free fermion case.\cite{Hunyadi04,er13} The later are related
to the singular dependence of the local Fermi momentum with the spatial position, at the front edge.
Our numerics are compatible with such  type of power law behavior of the period, although the exponent could be different.
In any case, the spatial period is much longer than the Fermi wave-length and these oscillations are reminiscent of some
soliton trains associated to shock waves in non-linear Luttinger liquids.\cite{baw06}
It is interesting to notice that in this interacting case the oscillations
take place {\it ahead} of the front which is in sharp contrast with the $\Delta=0$ case where the oscillation take place just behind
the leading edge of the front.\cite{Hunyadi04,er13}

\section{Entanglement entropy} 
\label{sec:EE}

The time evolution of the entanglement entropy was studied by Eisler {\it at al.}\cite{eip09} in
the particular case where $\Delta=0$, $m_0=\frac{1}{2}$
and $L=\infty$. There, it was shown that the entanglement associated to a left-right partition of the chain grows logarithmically with time.
Here we  describe the entanglement profiles obtained in finite chains for $m_0<\frac{1}{2}$ when varying the location of the boundary between the two subsystems.

\subsection{Steady state entanglement}

As for the DMRG method,  TEBD simulations are based on matrix-product states and they are all the more demanding to perform  as
the  entanglement entropy (EE) associated to left-right partitions of the chain is high.
On the other hand, global quenches often produce highly entangled states, with entanglement entropies which would scale as the
volume of the subsystem, as for thermal distributions. These situations are therefore difficult to simulate at long times.
The situation is quite different here, where the NESS entanglement entropy turn out to be rather low.
This is easy to understand for $\Delta=0$ since the NESS is in that case a boosted Fermi with exactly the same EE as the ground state.\footnote{A very long times,
much larger than the time for the front to reach the system boundary, even a Free fermion chain  will develop
a large EE, proportional to the subsystem size. This is not the regime we consider here, where the time $t$ is kept
smaller or equal than $L/v_{\rm max}$.}
Thanks to these relatively low entropies -- of order one  for a chain of length $L=80$ --
a good convergence is observed even with a rather small number of Schmidt eigenvalues  in the TEBD simulations (we use $\chi=100$ here).

\begin{figure}
\includegraphics[width=8cm]{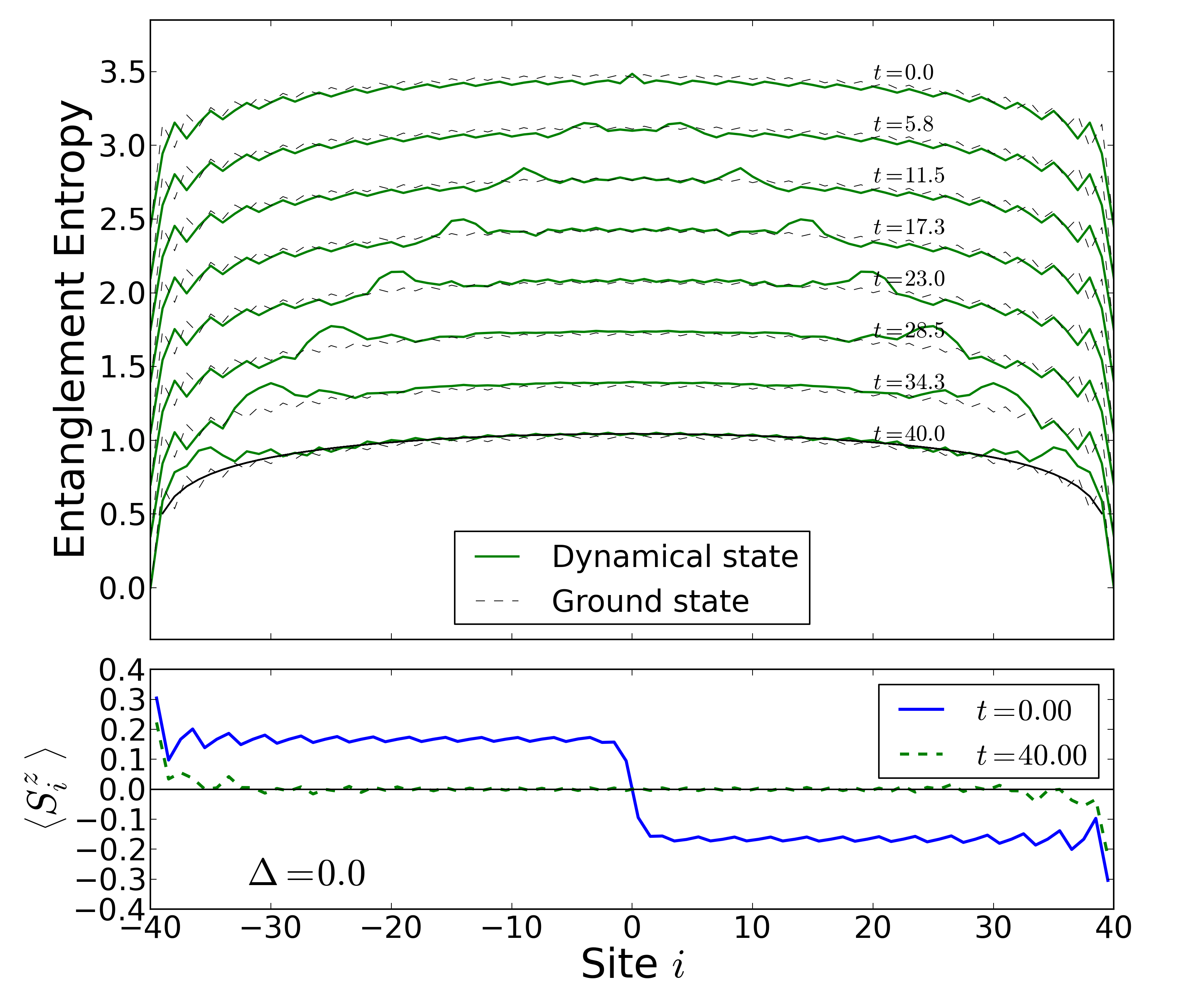}
\caption[99]{Entanglement entropy between the left part (sites $j<i$) and the right (sites $j\geq i$) as a function
 of the cut position $i$ and for different times (curves shifted by $0.35$ for clarity). Bottom panel: magnetization profile
 in the initial state and at $t=40$.
Dotted line: EE in the ground state of the chain, without external magnetic field. Full line: Eq.~\ref{eq:EE}. Parameters: $\Delta=0.0$ and initial magnetization $m_0=0.17$.
}
\label{fig:EEE-FFT}
\end{figure}

\begin{figure}
\includegraphics[width=8cm]{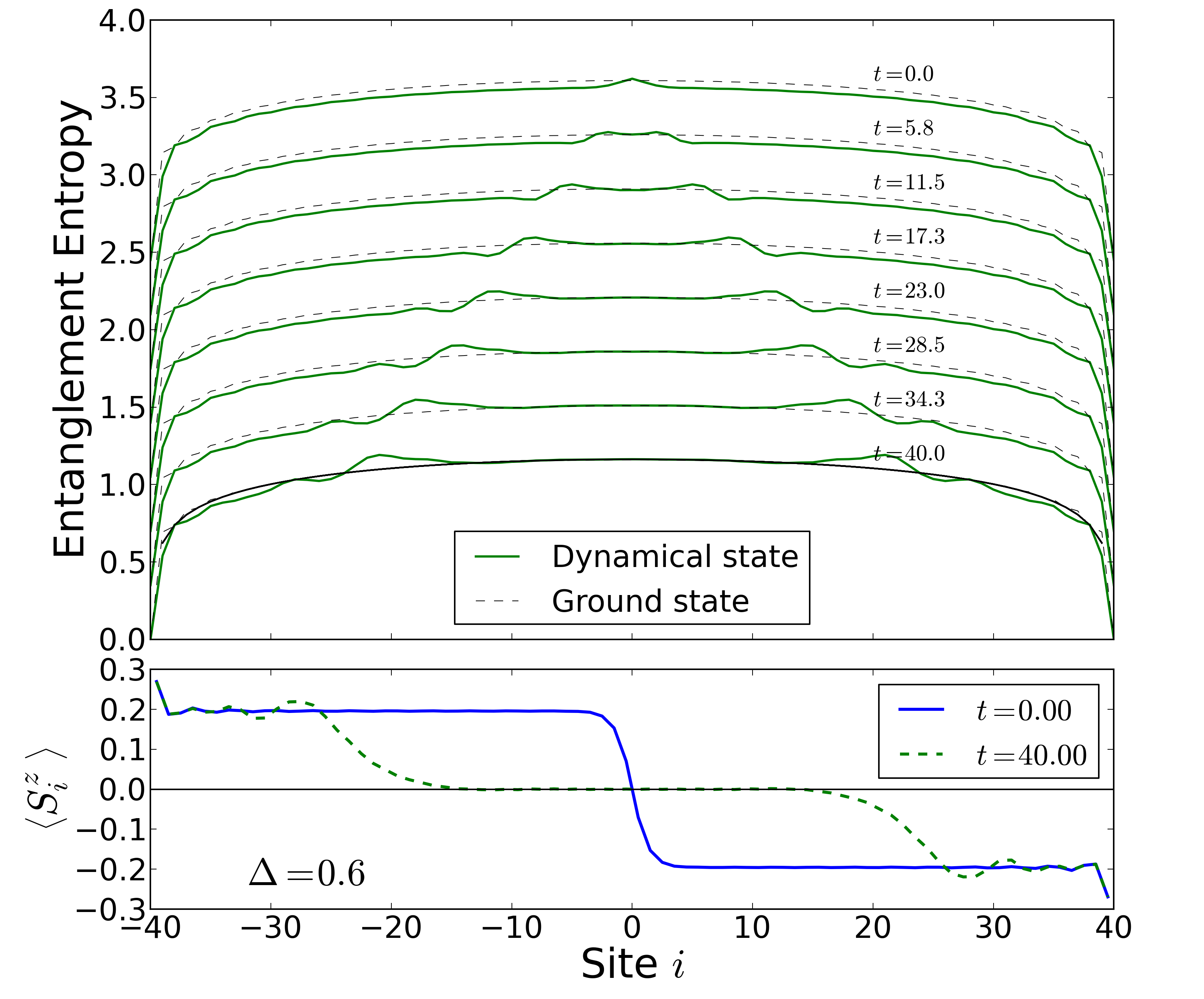}
\caption[99]{Same as Fig.~\ref{fig:EEE-FFT} for  $\Delta=0.6$ and $m_0=0.19$.
}
\label{fig:EEE-Delta06}
\end{figure}

Figs.~\ref{fig:EEE-FFT} and \ref{fig:EEE-Delta06} show that the entropy stays relatively small during the whole evolution,\footnote{If we had a weaker or stronger  bond in the middle of the chain the situation would be very different: an incoming wave (fermion) would be partly reflected and partly transmitted.
Each such event would contribute by a finite amount to the left-right entropy and the finite current would
imply an EE increasing  linearly in time.}
and hardly exceeds the ground state value (dotted lines and Eq.~\ref{eq:EE}). The only situation where the EE is above the ground state value is
when the cut is inside the region of a front. The front location is indeed clearly visible on the EE plots and the space-time picture shows
a  characteristic ``light cone'' shape.

The NESS region appears to have an entropy profile which is very close to that of the ground state.
The later is well described by conformal field theory and
the leading term in the entropy of a segment of length $l$ in  an open  critical chain (central charge $c=1$) of length $L$ is:\cite{entropy}
\begin{equation}
 S(l)=\frac{1}{6}\ln\left[L \sin\left(\frac{\pi l}{L}\right)\right]+C
 \label{eq:EE}
\end{equation}
($C$ is a non-universal constant).
The agreement with the numerics in the NESS region suggests that the NESS is not a thermal-like state with extensive entropy, but is instead entropically close
to some low-energy critical state. This is indeed the case for $\Delta=0$ case since
the NESS is a boosted Fermi sea whose entanglement profile is identical to the Fermi sea ``at rest'' (ground state).
In the bosonization approach\cite{lm10} the NESS 
can be described by adding a classical ``twist'' $S^+(x,t) \to S^+(x,t)\exp{(i h x / v )}$ ($v$ is the velocity)
and therefore shares the same entanglement profile as the ground state.

\subsection{Entanglement between the left- and right- moving fronts}
\label{ssec:step}

\begin{figure}
\includegraphics[width=8cm]{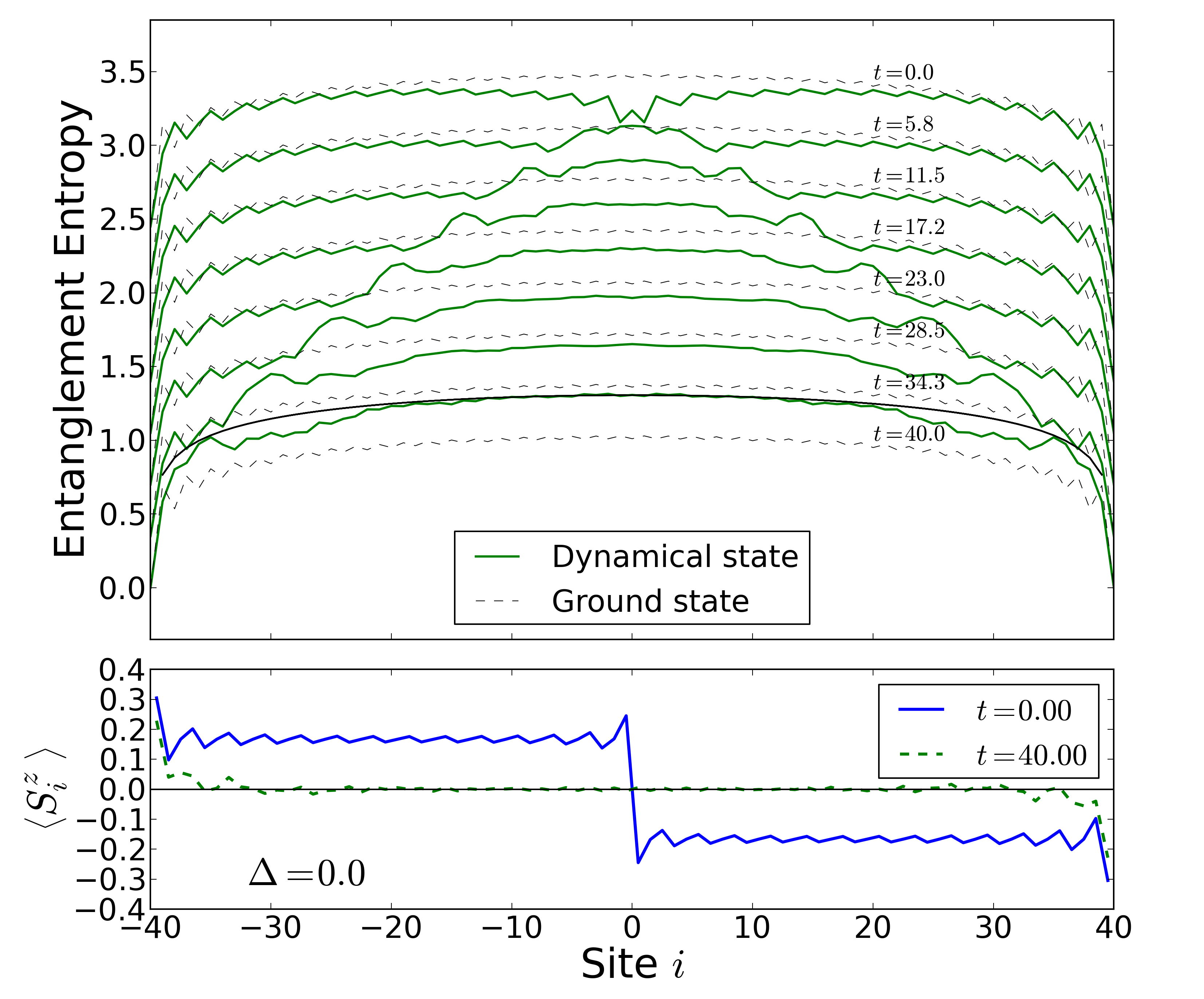}
\caption[99]{Same as Fig.~\ref{fig:EEE-FFT} but with
an initial state constructed from a ``step'' magnetic field profile (Eq.~\ref{eq:step}).
Parameters: $\Delta=0.0$ and initial magnetization $m_0=0.17$.}
\label{fig:EEE-FFS}
\end{figure}

Here we discuss the influence of the initial conditions on the entanglement entropy profile.
More specifically, we compare the two following initial conditions: the smooth ``$\tanh$'' profile of Eq.~\ref{eq:tanh}
and a ``step'' profile associated to the following Hamiltonian:
\begin{equation}
H_{t<0}=H + h \left(- \sum_{i=-L/2}^{-1} S^z_i + \sum_{i=0 }^{L/2-1} S^z_i\right)
\label{eq:step}
\end{equation}
The two situations lead to the same NESS at long times, but the entanglement profiles are different in both cases.
The EE profiles associated to this step initial condition is plotted 
in Fig.~\ref{fig:EEE-FFS} in the $\Delta=0$ case.
These result should be compared with Fig.~\ref{fig:EEE-FFT}: in the ``step'' case the EE of the NESS region is shifted
by some constant $S_0$ of order one. This shift naturally interpreted as a contribution from the entanglement between the left front with the right front.
Although this does not physically affect the NESS, it is numerically advantageous to start from a smooth ($\tanh$) magnetic field in order to minimize the EE between
the left- and right-moving excitations that form the fronts.\footnote{For instance, doing so can reduces by $0.3$ the EE and can thus reduce by a factor $e^{-0.3}=0.74$ the
required number $\chi$ states to be kept in the TEBD simulation, and a factor $0.74^3=0.41$ in the simulation time.}

Below we provide below an intuitive explanation of this phenomenon.
In the limit where the initial magnetic field varies smoothly at the lattice spacing scale, one may consider that the fermions form locally a Fermi sea.
In that situation, each occupied state has a momentum and velocity and the classical picture predicts that the particles will flow to the left or two the right
at constant velocity. In this picture, there is no particular entanglement associated to the fronts. Now consider a weak but sharp magnetic field.
One can view this as a perturbation of the homogeneous Fermi sea. But since the spatial dependence of the perturbation is sharp in real space, it contains many Fourier components.
As a consequence, the initial state contains some excited particles which are in a {\it linear combination of different momentum states}. Since the perturbation is weak, these
will be close to $\pm k_F=\pm \pi/2$. During the evolution
the wave function of these excitations will split into a left- and right- moving parts. Naturally, these two parts are entangled (would be as high as $\log(2)$ for
$\frac{1}{\sqrt{2}}\left(c^\dagger_{p} + c^\dagger_{-p}\right)$). In this picture the two fronts turn out to be entangled
due to the presence of wave-packets with Fourier components at $p\sim \pm\pi/2$ which are initially created at the origin by the magnetic field step.

\subsection{Classical picture, very long times and extensive entanglement entropy}
\label{ssec:long_times}

Although we are mostly interested in the dynamics before the fronts reach the ends of the chain, we present here some data concerning 
the evolution at times $t$ much larger than $L/v_{\rm max}$. Doing so is numerically not possible with TEBD and we therefore
focus on the free fermion chain at $\Delta=0$.

\begin{figure}
\includegraphics[width=7cm]{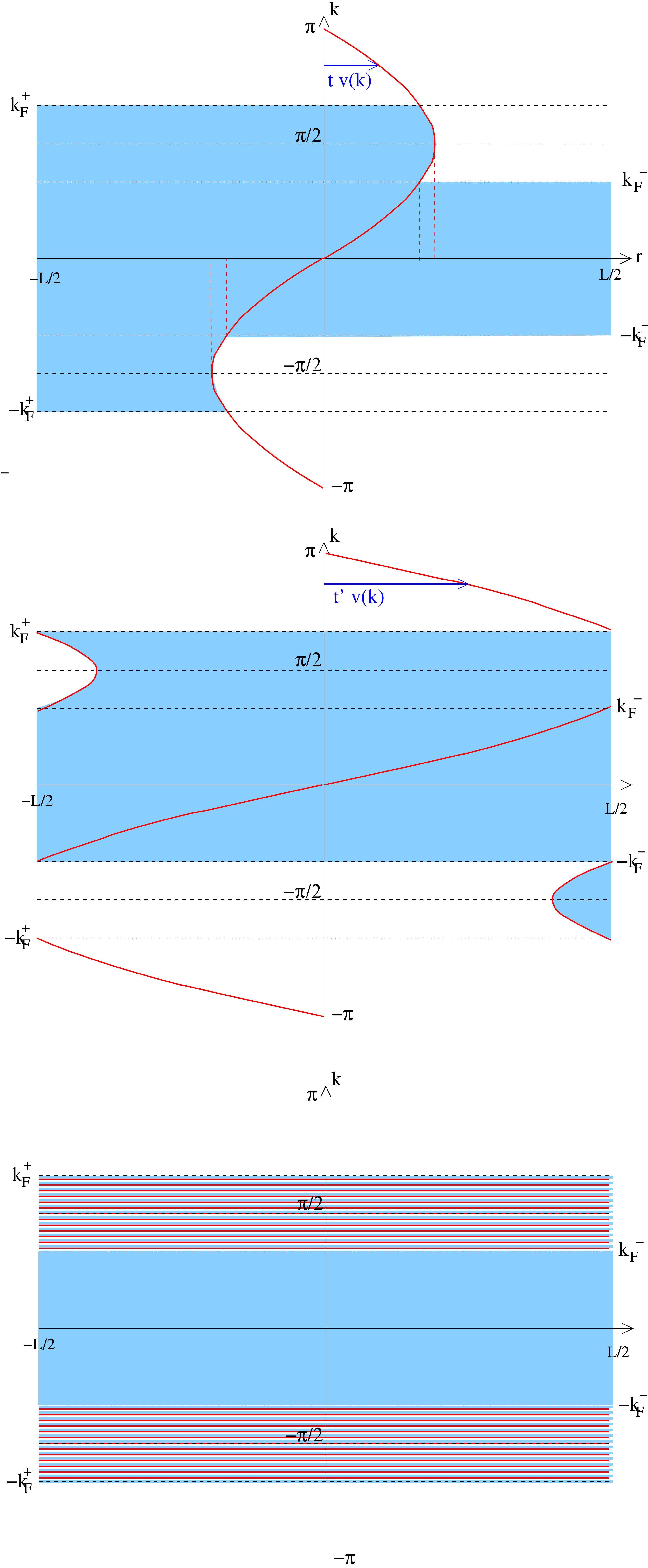}
\caption[99]{(color online) Classical evolution of the density $n(p,r,t)$.
The occupied region of phase space, where $n(p,r,t)=1$ is colored in blue (zero otherwise). The red curve
represents the position of a particle with momentum $p$ located initially at $r=0$.
For $k_F^-\leq|p|\leq k_F^+$, this curve separates the occupied ($n=1$) from the empty region ($n=0$) of phase space.
Top : Short time $t'$. The locations of  front of the fronts and back of the fronts are indicated by dashed red lines.
Center: time $t'$ after the first bounce at the boundaries. Bottom: In the long time limit $L\ll t$
the red curve spans the interval $[-L/2,L/2]$ many times and the 
the striped area therefore corresponds
to a alternation of thin occupied and empty regions, with an average density $n(p,r,t=\infty)=\frac{1}{2}$ for $k_F^-\leq|p|\leq k_F^+$.}
\label{fig:semic}
\end{figure}

\begin{figure}
\includegraphics[width=8cm]{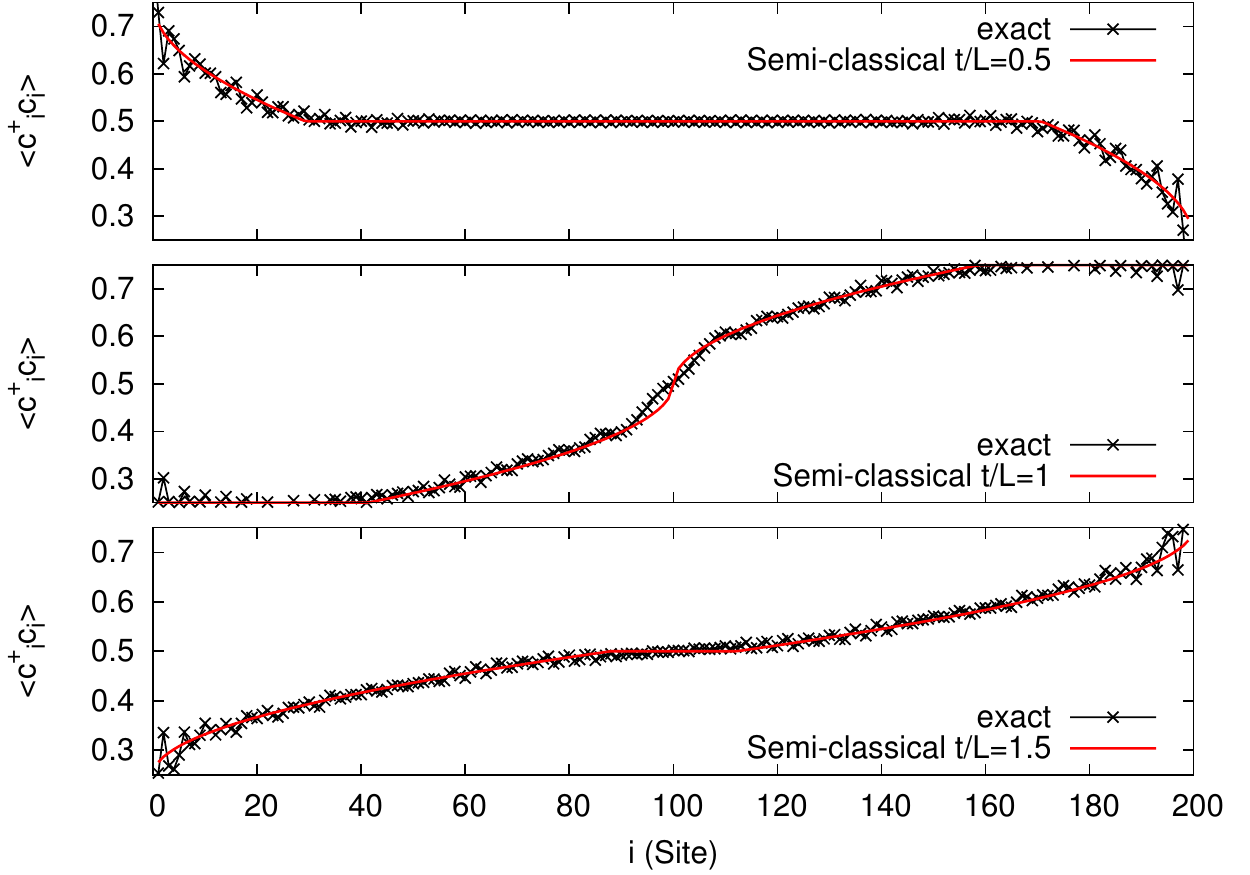}
\includegraphics[width=8cm]{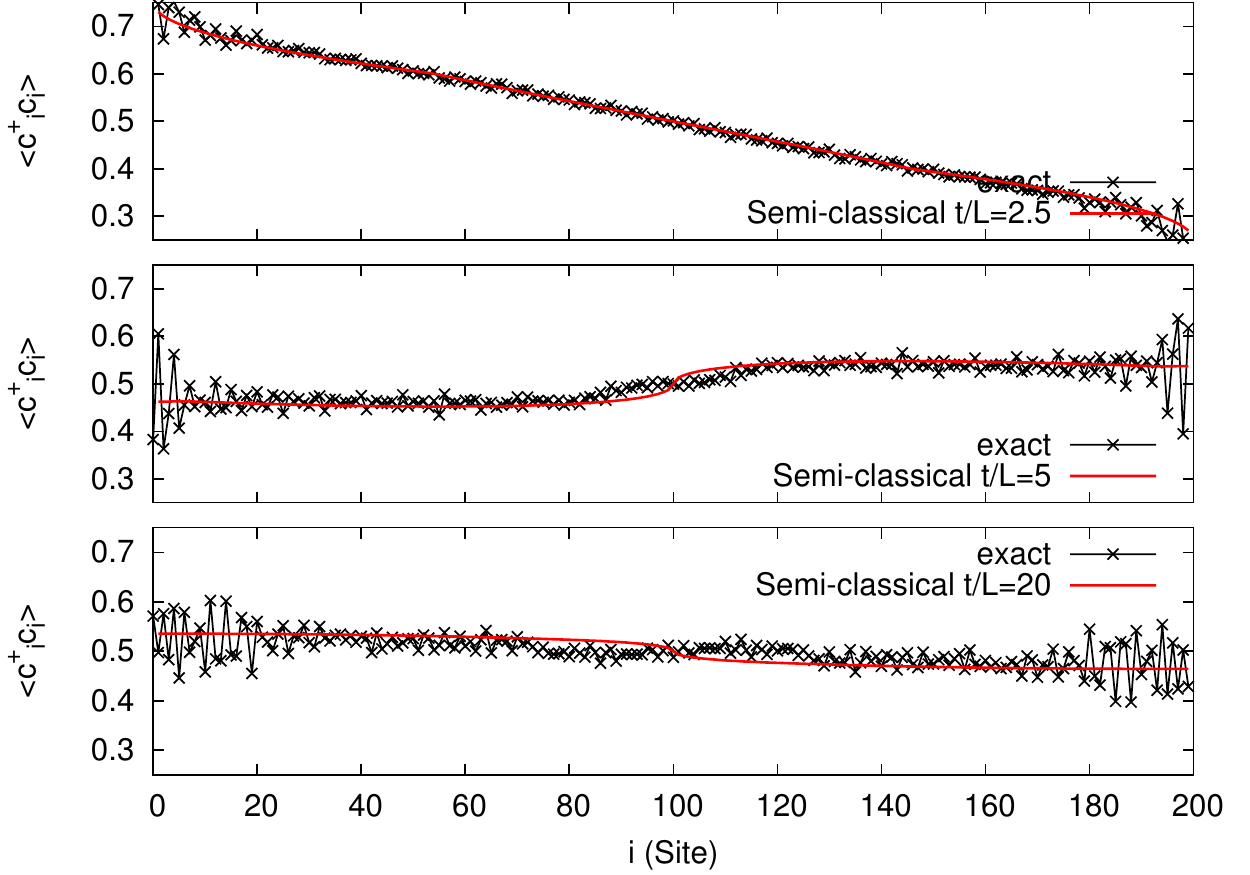}
\caption[99]{Density profile for $\Delta=0$, $L=200$  and $h_0=\sqrt{2}/2$ (corresponding to densities
0.25 and 0.75 at $t=0$). Crosses: exact result. Red: classical result (thermodynamic limit).}
\label{fig:lonTimeProfile}
\end{figure}

A classical (also called hydrodynamical), description was introduced in 
Ref.~\onlinecite{antal08}. In this approximation
the system is characterized by the density $n(p,r,t)$ of particles having a well-defined position $r$ {\it and} momentum $p$ at time $t$.
At the initial time, this function describes two Fermi seas at different densities for $r>0$ and $r<0$:
\begin{eqnarray}
n(p,r<0,t=0)&=&\left\{\begin{array}{c}
              1\;{\rm for}\;    |p|\leq k^+_F \\           
              0\;{\rm for}\;    |p|>    k^+_F
             \end{array}\right. \nonumber \\
 n(p,r>0,t=0)&=&\left\{\begin{array}{c}
              1\;{\rm for}\;    |p|\leq k^-_F \\           
              0\;{\rm for}\;    |p|>    k^-_F
             \end{array}\right.        \label{eq:init}     
\end{eqnarray}
with the Fermi momenta $k_F^\pm$ given by Eq.~\ref{eq:kfpm}.
Then,  each particle (fermion) with momentum $p$ propagates ballistically at velocity $v_p=\sin(p)$:
\begin{equation}
 \frac{\partial}{\partial t} n(p,r,t)=-v_k \frac{\partial}{\partial r} n(p,r,t)
\end{equation}
For the initial condition of Eq.~\ref{eq:init}, the time evolution amounts to follow  the occupied ($n=1$) and
empty regions ($n=0$) of phase space ($r$ and $p$) according to the free particle propagation.
This is schematically represented in Fig.~\ref{fig:semic}.
The total density (or magnetization) $n(r,t)$ at a given point $r$ is obtained
by integrating $n(p,r,t)$ over momenta: $n(r,t)=\int_{-\pi}^{\pi} n(p,r,t) dp/(2\pi)$.
For $1\ll t \ll L$ this approximation reproduces the exact shape of the front in the limit of long times:
\begin{equation}
n(r,t)=\left\{\begin{array}{cl}
	\frac{1}{2} & {\rm for}\;0 \leq r/t \leq \sin(k_F^+) \\
        \frac{\arccos(r/t)}{\pi}+\frac{1}{2}-m_0 &{\rm for}\; \sin(k_F^+) \leq r/t \leq 1 \\
	 \frac{1}{2}-m_0 &{\rm for}\; 1 \leq r/t 
       \end{array}\right.
\end{equation}
When a particle reaches the end of the chain we then assume that the momentum (and thus velocity) is simply reversed
(see bottom of Fig.~\ref{fig:semic}). This approach
leads to the results show in Fig.~\ref{fig:lonTimeProfile}. The comparison between the exact result and the classical/hydrodynamical approximation
shows that the agreement only slightly deteriorates at long times. 

\begin{figure}
\includegraphics[width=6.5cm]{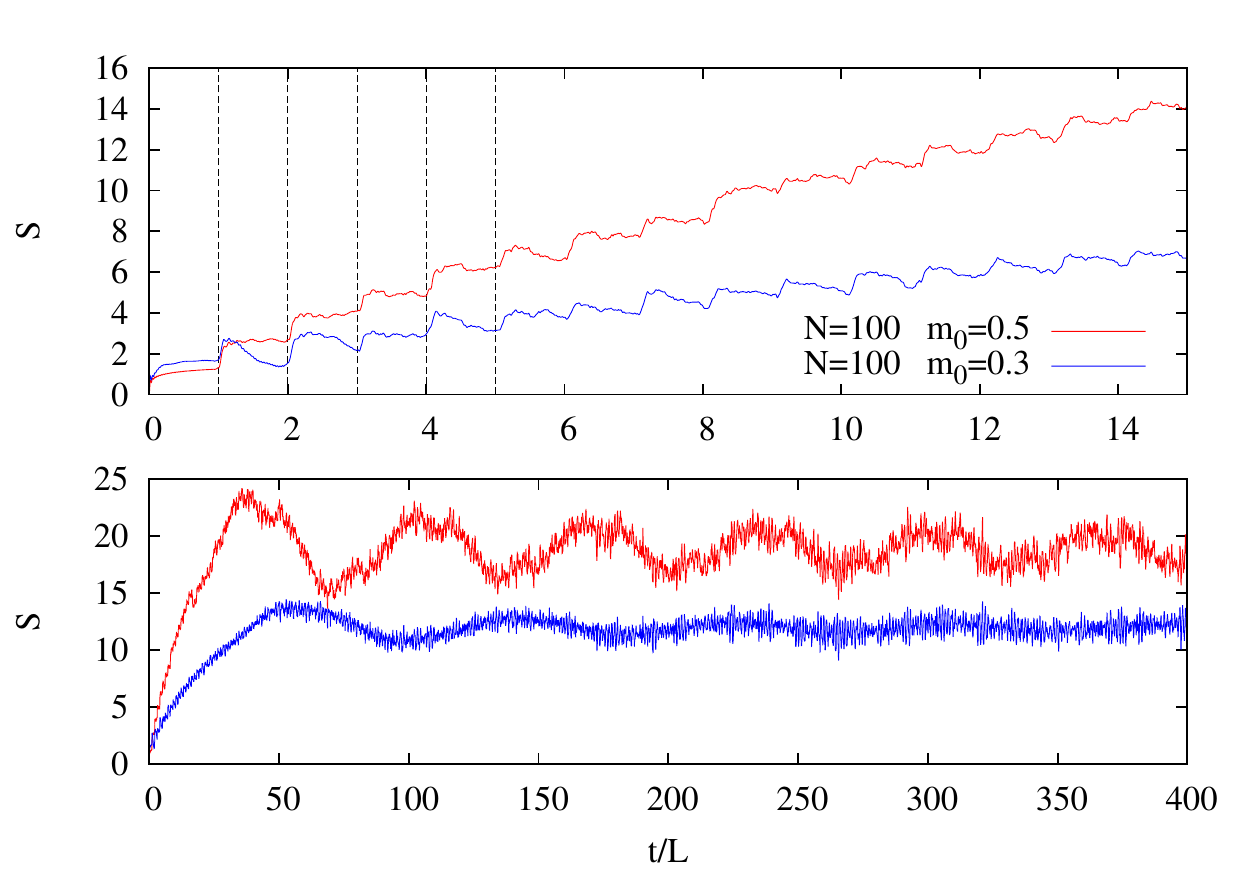}
\caption[99]{Entanglement entropy between the two half of the chain as a function of time (scale by $L$).
The vertical lines correspond to the times where the two fronts cross each other in the center of the chain.
The limit $t/L\ll 1$ for $m_0=\frac{1}{2}$ corresponds to the logarithmic increase of the entropy studied in Ref.~\onlinecite{eip09}.
}
\label{fig:lonTimeEE}
\end{figure}

With a Fermi velocity equal to $v=1$, the two fronts cross at  $t=L$, $2L$, $3L$, etc.
The Fig.~\ref{fig:lonTimeEE} shows a rapid increase of entropy each time the front cross (vertical lines). After a number  crossings proportional to the system size,
the magnetization slowly equilibrate to $\langle S^z\rangle=0$ (data not shown) and the entanglement becomes {\it extensive}
(see also Ref.~\cite{zangara} for an exact diagonalization study of the long time limit in small chains). 

We observed in the numerics that for $t\gg L$ (many bounces) the exact fermionic
correlations become diagonal in momentum space (when measured sufficiently far from the system boundaries) :
\begin{equation}
 \langle c^\dag_{p'} c_p\rangle_{t\gg L} \sim \delta(p-p') n(p).
 \label{eq:np}
\end{equation}
This diagonal form is equivalent to the translation invariance of $\langle c^\dag_i c_j\rangle_{t\gg L}$ and can be intuitively understood from the fact
that the system loses the ``memory'' of the initial front location (see also Sec.~\ref{sec:free}).
$n(p)$ is the initial occupation number $\langle c^\dag_p c_p\rangle_{t=0}$ defined on the whole chain (average of the left and right reservoir contributions).
This conserved quantity is the sum of two Fermi distributions with two different
Fermi momenta $k_F^+$ and $k_F^-$. Assuming $k^+_F>k_F^-\geq0$ we have:
\begin{equation}
 \langle c^\dag_p c_p\rangle_{t\gg L}=
 \left\{\begin{array}{cl}
0&{\rm for}\;    |p|>k_F^+ \\
\frac{1}{2}&{\rm for}\;    k_F^+>|p|>k_F^-\\
1&{\rm for}\;    k_F^->|p|
\end{array}\right.
\end{equation}
In the infinite time limit this distribution is also naturally obtained from  classical picture (bottom of Fig.~\ref{fig:semic}).

The reduced density matrix and the entanglement entropy of a segment can be constructed entirely from its two-point correlations.\cite{peschel}
The fact that some of the fermion modes are {\it partially} occupied ($n(p)=1/2$ for $k_F^+>|p|>k_F^-$) naturally leads to an
 {\it extensive} entanglement entropy. Indeed each of these partially occupied modes contribute by an amount $\log(2)$ to the Von Neumann entropy.
 So, from a simple counting of the number of these modes
 we can expect the entropy of a segment of length $l$ to scale as $\sim l \log(2) |k_F^+-k_F^-|/\pi=2 l\log(2) |m_0| $.
 This is consistent with our numerics as well as with a direct and rigorous calculation.\cite{jms}
The large entropies observed at long times in Fig.~\ref{fig:lonTimeEE} can thus be explained by the emergence of a Fermi distribution
with partially occupied modes in the interval $k_F^+>|p|>k_F^-$. 
This long time limit is a particular realization of the dephasing phenomenon discussed in Ref.~\onlinecite{bs08}.

\section{Current dynamics and conductance}
\label{sec:current}

\begin{figure}
\includegraphics[width=9cm]{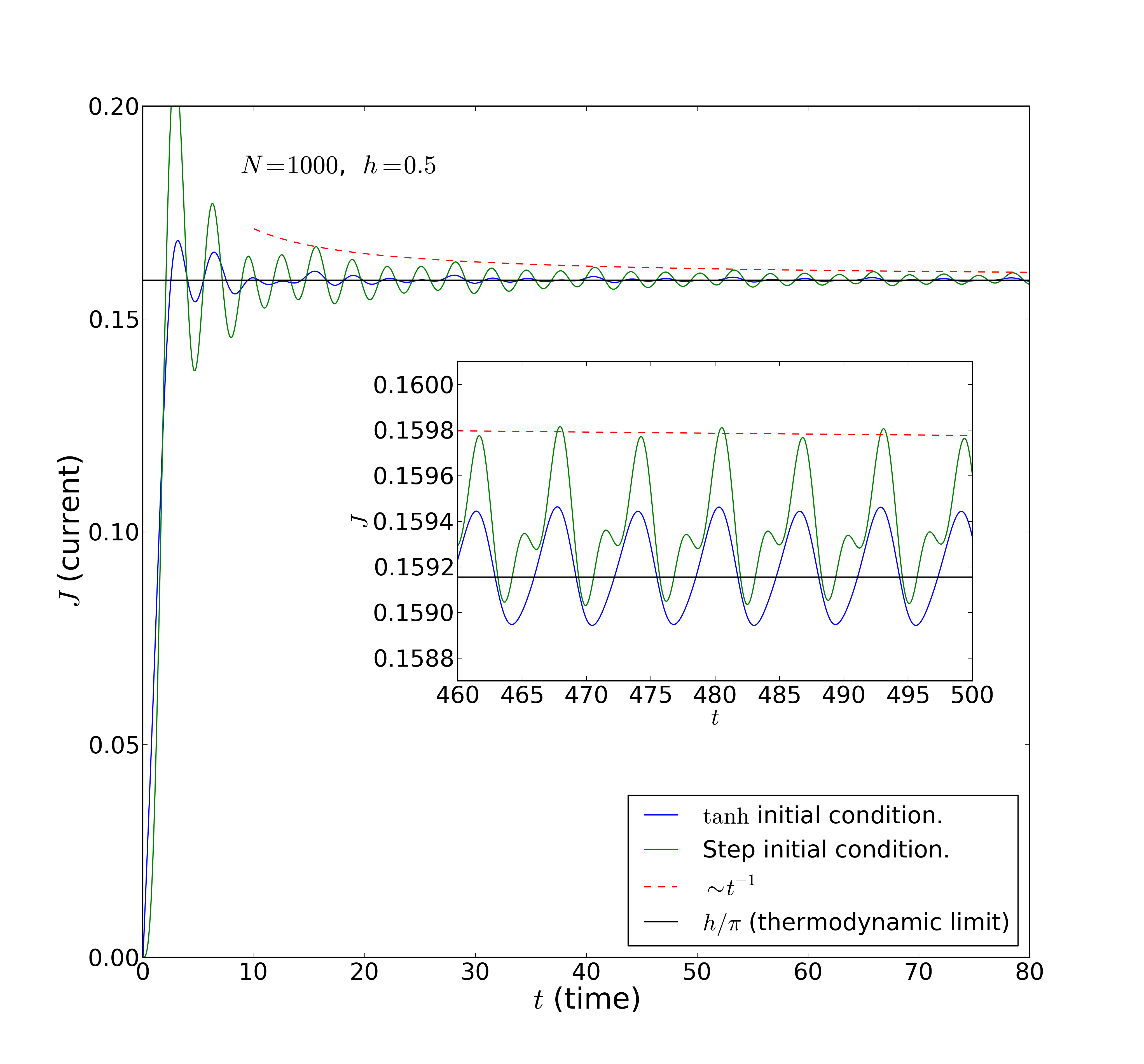}
\caption[99]{(Color online)
Time evolution of the spin current $J$ measured in the center of a chain of length $L=1000$  and for $\Delta=0$ (free fermions).
Green curve: step initial magnetic field. Blue: $\tanh$ initial magnetic field.}
\label{fig:currentFF}
\end{figure}

We analyze here how the current reaches a stationary value.
We first focus on the free Fermion case ($\Delta=0$), for which very long chains and long times can easily be studied.
In Fig.~\ref{fig:currentFF} the current $J(t)=\Im \langle S^+_0 S^-_1\rangle$ measured in the center of a long chain is plotted as a function of $t$.
We compare two initial conditions: the step (green) and $\tanh$ (blue) initial magnetic field.
In both cases the current rapidly reaches a quasi stationary regime with small amplitude residual oscillations ($\mathcal O(1/t)$)
which have a period $t=2\pi$.
These oscillations have been previously observed and analyzed in Ref.~\onlinecite{ecj12}.
This quasi stationary regime is attained at relatively short times, $t\simeq 30-40$.
When the oscillations are averaged out, the value  of the current is then close
to $h/\pi$, the current carried by the boosted Fermi sea (black horizontal line). For the $\tanh$  initial conditions the oscillations turn out to be slightly smaller
and the average current is closer to the thermodynamic value $h/\pi$.

\begin{figure}
   \includegraphics[width=9cm]{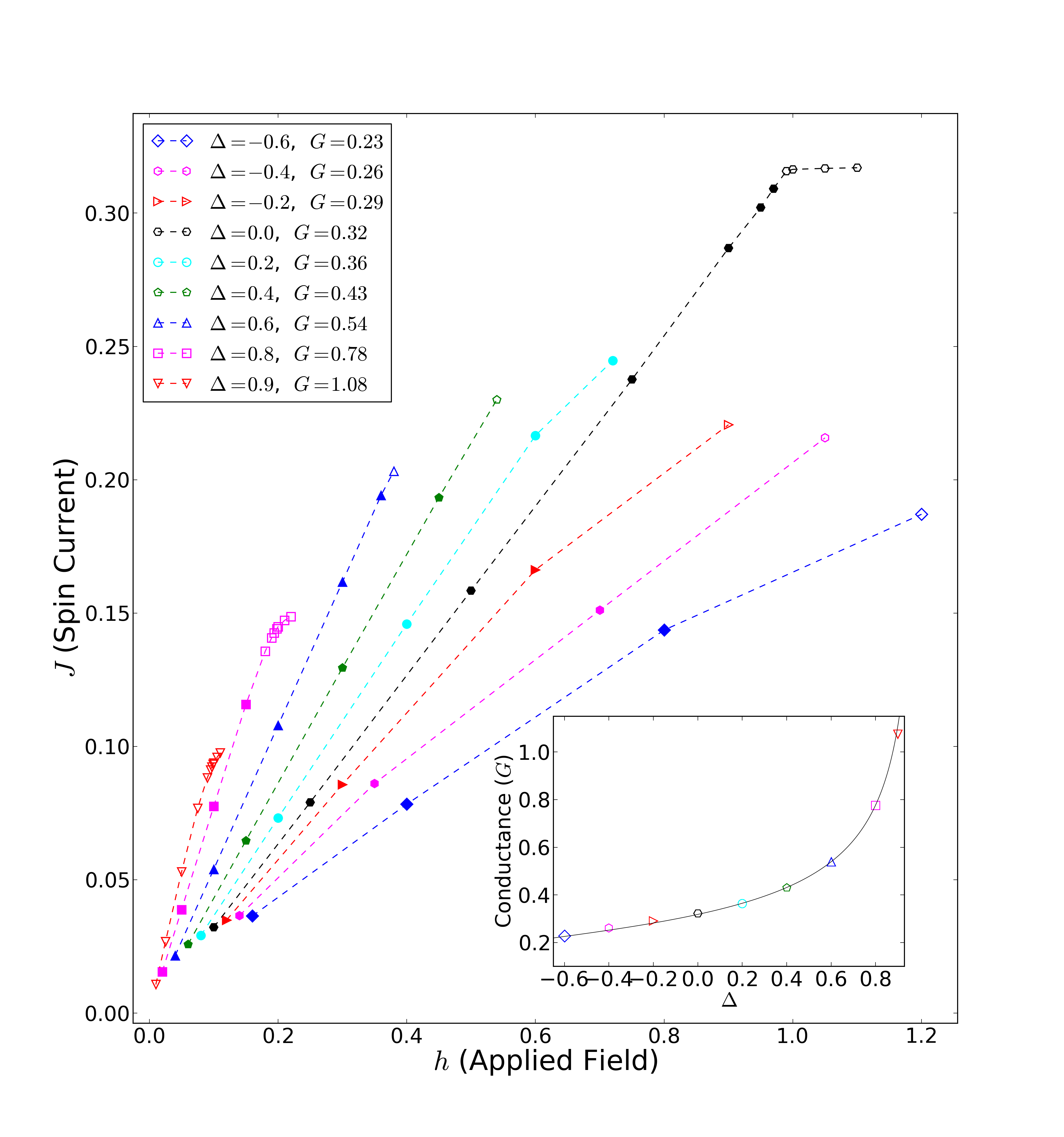}
   \caption[99]{Spin current $J=Im \langle S^+_{0} S_{1} \rangle$ measured in the center of the chain in the steady state regime.
   Inset: The slope $G=dJ/dh_{|h=0}$ (conductance) is plotted as a function of $\Delta$, and matches the Tomonaga-Luttinger liquid prediction (Eq.~\ref{eq:GLL}).
}
\label{fig:conductance}
\end{figure}

From this we conclude that averaging over a few oscillations the current for $t\simeq 40$ is a legitimate way to estimate numerically the value of the stationary current.
This is the procedure used to obtain the results shown in Fig.~\ref{fig:conductance}.

Although the current is not strictly linear in $h$ (except in the free fermion case for $h<1$), we observe
an extended linear regime. The slope $G=dJ/dh_{|h=0}$ is the conductance of the system, and
the values obtained numerically matches the Tomonaga-Luttinger (TL) liquid prediction (inset of Fig.~\ref{fig:conductance}):\cite{fg96}
\begin{equation}
 G=\frac{K}{\pi}
 \label{eq:GLL}
\end{equation}
in units where the ``electric charge'' $e$ and $\hbar$ are set to unity. This result is in agreement Ref.~\onlinecite{ecj12}.
It is however interesting to note that the TL conductance is obtained here in an isolated quantum, and therefore without any dissipation.
This may sound counter intuitive since a finite conductance $G$ is naturally associated to a dissipated power $P=J h$ ($h$ is the chemical potential difference).
This is made possible by the fact that the inhomogeneous external magnetic field $h$ is switched off during the evolution.
This way, a current can flow from the high density  reservoir to the low density reservoir while keeping constant the energy.

\section{Correlations in the stationary state}
\label{sec:correl}

\begin{figure}
 \includegraphics[width=9cm]{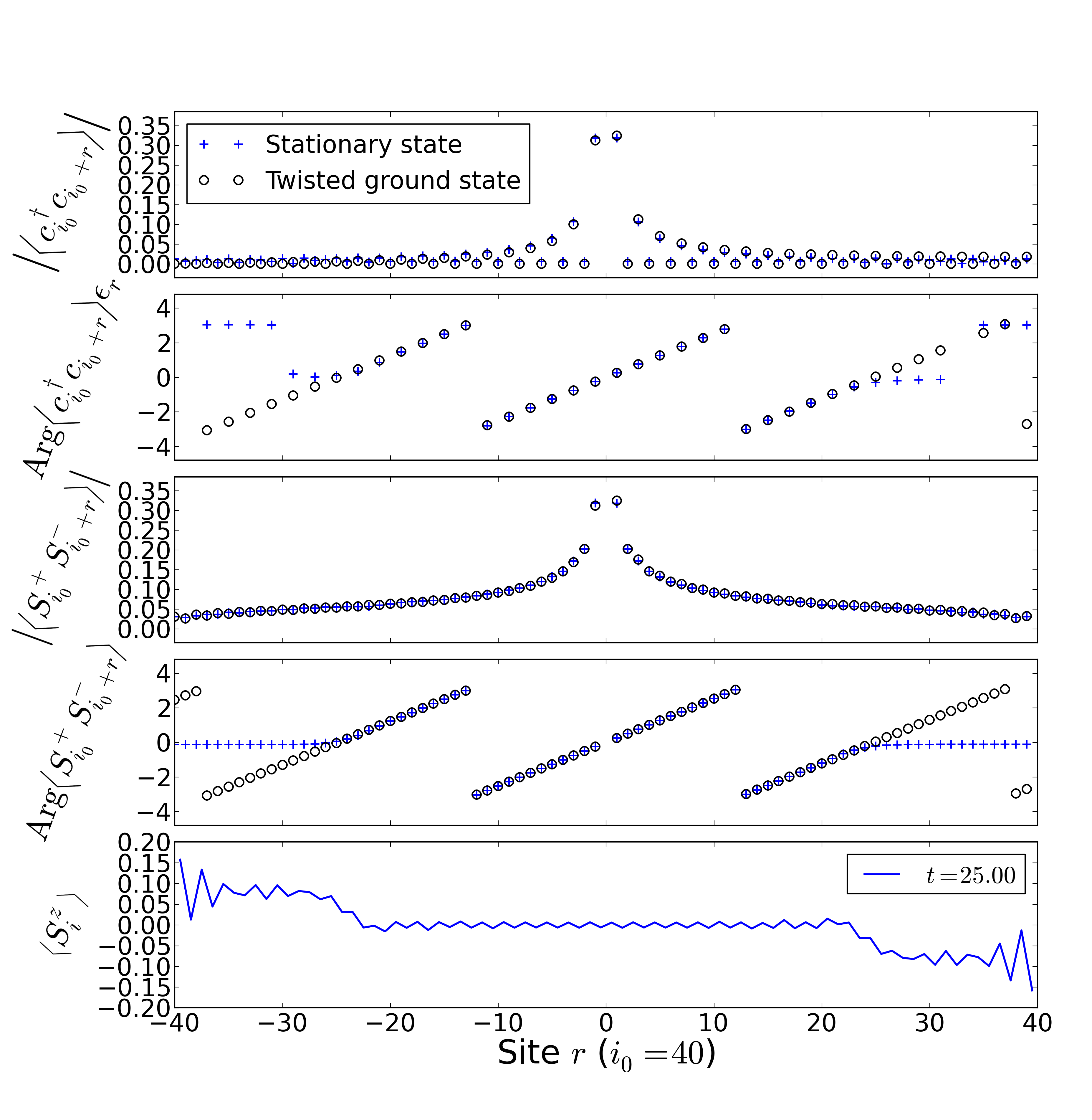}
  \caption{(Color online) Crosses: various correlations
  measured at time $t=40$ in a chain of length $80$ for $\Delta=0$ and $h/h_{\rm sat}=0.25$ (TEBD with $\chi=100$ states), and $m_0\simeq 0.08$.
    From top to bottom: i) Modulus of $\langle c^\dagger_{i0} c_{i0+r}\rangle$, ii) Argument of $\langle c^\dagger_{i0} c_{i0+r}\rangle$
     iii)  Modulus of $\langle S^+_{i0} S^-_{i0+r}\rangle$ iv) Argument of $\langle S^+_{i0} S^-_{i0+r}\rangle$. v) magnetization.
    For i-iv, the circles represent the same correlator measured in the zero-magnetization ground state and multiplied by a phase factor $e^{i\theta r}$.
    $\theta$ is determined numerically by a fit over 5 sites in the middle of chain (from 41 to 45). } \label{fig:CorrelDelta0}
 \end{figure}

\begin{figure}
 \includegraphics[width=9cm]{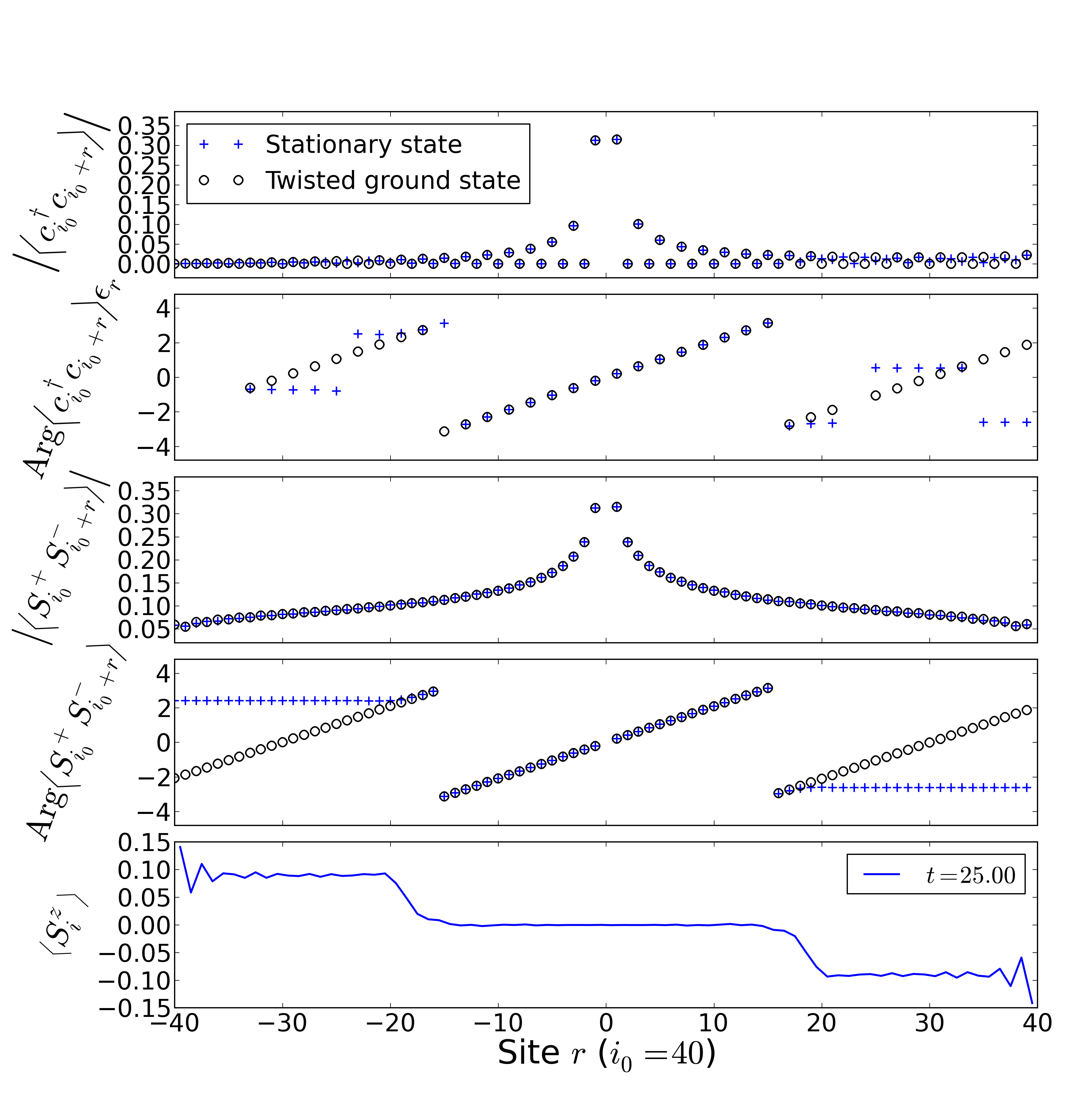}
  \caption{(Color online) Same as Fig.~\ref{fig:CorrelDelta0} for $\Delta=0.4$, $h/h_{\rm sat}=0.25$.}\label{fig:CorrelDelta04}
 \end{figure}

\begin{figure}
 \includegraphics[width=9cm]{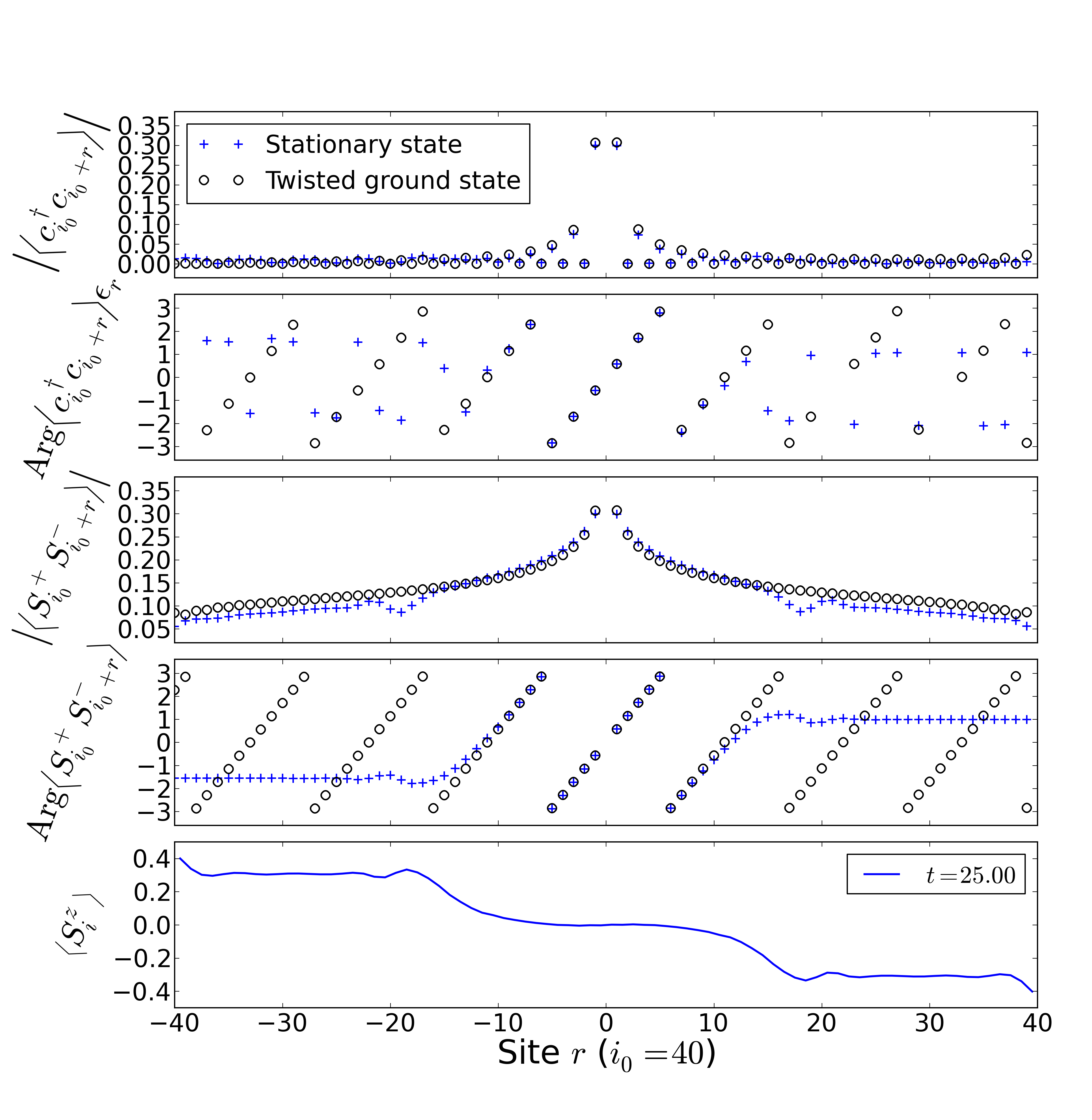}
  \caption{(Color online) Same as Fig.~\ref{fig:CorrelDelta0} for $\Delta=0.6$, $h/h_{\rm sat}=0.75$.}\label{fig:CorrelDelta06}
 \end{figure}

\begin{figure}
 \includegraphics[width=9cm]{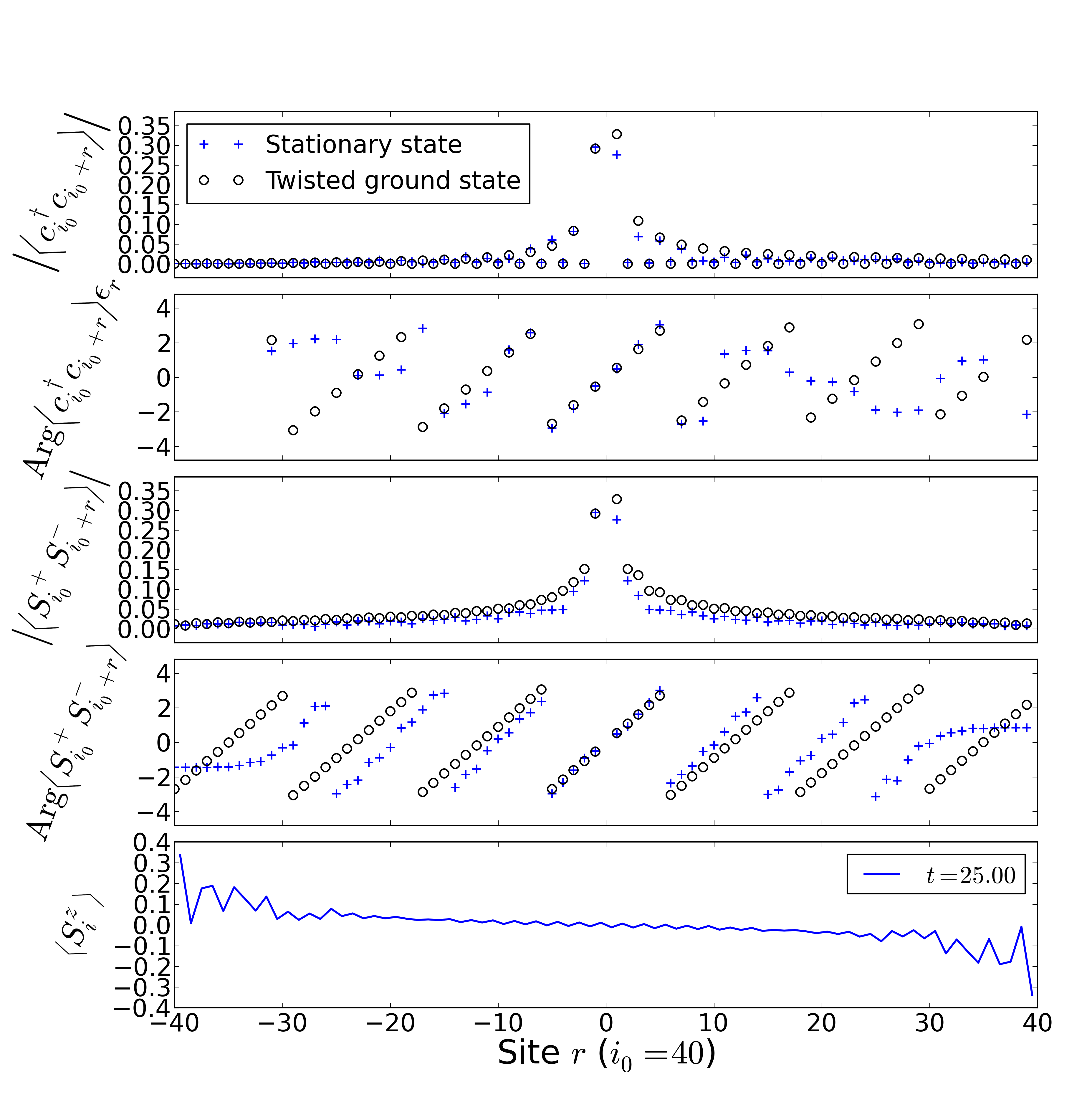}
  \caption{(Color online) Same as Fig.~\ref{fig:CorrelDelta0} for $\Delta=-0.6$, $h/h_{\rm sat}=0.6$.}\label{fig:CorrelDelta-06}
 \end{figure}

In this section we discuss the two-point correlations in the stationary region.
The numerical results are summarized in Figs.~\ref{fig:CorrelDelta0}, \ref{fig:CorrelDelta04},\ref{fig:CorrelDelta06} and \ref{fig:CorrelDelta-06}.

{\it Non-interacting case}  ---. In the thermodynamic limit and at long times the stationary state is known to be a ``boosted'' Fermi sea for $\Delta=0$. 
As discussed in Sec.~\ref{sec:ff}, the Eqs.~\ref{eq:SpSmtwist} hold exactly in that case, with a twist angle
$\theta$ (Fermi momentum shift) given by $\theta=\pi m_0$.
The magnetization profile (bottom panel of Fig.~\ref{fig:CorrelDelta0})  shows that the fronts have not reached the ends of the chain at $t=25$ (for this value of $h$ the velocities
are $v_{\rm max}=1$ and $v_{\rm min}=\cos(\pi m_0) \simeq 0.968$).
The magnetization shows some oscillations that are the spatial
counter-part of the temporal oscillations displayed in Fig.~\ref{fig:currentFF}.
The oscillations have a spatial period of two sites (the center of the chain is close to half-filled).
When the front has reached the end of the chain, the amplitude of these oscillations is $\mathcal{O}(1/L)$. This is a finite-size effect but
it remains sizable for a chain of length $80$.
At the same time, Fig.~\ref{fig:CorrelDelta0} indicates that the two-point correlations (fermionic as well as spin-spin)
almost perfectly match Eqs.~\ref{eq:SpSmtwist} in the central region. 
What we learn here is that the two-point correlation functions converge relatively rapidly to their asymptotic form, even
for relatively short times and small chains.

{\it Interaction and weak current} ---. Next, in Fig.~\ref{fig:CorrelDelta04} we investigate a situation with moderate interactions ($\Delta=0.4$) and a weak bias $h/h_{\rm sat}=0.25$.
Compared to the free fermion case, the fronts propagate at slightly lower velocities.  Still, a large central region
exhibit a constant  magnetization $\langle S^z \rangle \simeq 0$ and the $[-15,15]$ can be considered as almost stationary.
We note that the spatial oscillations observed for $\langle S^z\rangle$ in the free fermion case are practically
invisible here.\footnote{As discussed in Ref.~\onlinecite{hf04},
a critical  open spin chain in an external magnetic field shows
some oscillations of $\langle S^z_i\rangle$  
which decay algebraically with the distance to the boundary.
This exponent is equal to the Luttinger parameter $K$ and is larger (hence faster decay) in presence of ferromagnetic interactions than for free fermions.
A similar phenomenon is presumably at play here and explains why oscillations are much smaller for $\Delta>0$.}

When inspecting the $\langle S^+_i S^-_j\rangle$ and $\langle c^+_i c_j\rangle$ correlations we see that, in the region $[-15,15]$ corresponding
to an almost flat magnetization, the moduli of the correlations match those of the ground state.
As for the complex argument of these two-point functions, it is a linear function of the distance $r$ between the two points. The slope of this argument
is used to determine numerically the twist angle $\theta$.

The agreement with Eqs.~\ref{eq:SpSmtwist} was expected for $\Delta=0$ (at least for long times in long chains),
but the agreement here in presence of interactions (Fig.~\ref{fig:CorrelDelta04}) was a priori not granted.
In fact, as discussed in Sec.~\ref{sec:boso} this is the bosonization prediction\cite{lm10} for the NESS.
A simple bosonization calculation is  not accurate to describe quantitatively the ground state correlations $\langle S^+_{i_0} S^-_{i_0+r}\rangle_{\rm gs}$
{\it at short distances} (of the order of one lattice spacing).
Similarly, this approach is a priori not expected to be precise to describe the NESS correlations $\la S^+_{i_0} S^-_{i_0+r}\ra_{\rm NESS}$ when $|r|\sim \mathcal{O}(1)$. But still, we find
that the ratio $\la S^+_{i_0} S^-_{i_0+r}\ra_{\rm NESS} / \la S^+_{i_0} S^-_{i_0+r}\ra_{\rm gs}$ is remarkably close to a pure phase factor $e^{ir\theta}$ (Eq.~\ref{eq:SpSmtwist}).
Several aspects of {\it interaction} quenches in the TL model have been studied\cite{cazalilla06,ic09,TLquenches,karrasch12} 
but the present setup (``Antal's quench'') is particularly useful if one is interested in comparing the lattice results to the bosonization prediction for the NESS.

{\it Interaction and strong current} ---.
At higher currents, one starts to observe some small deviations from Eq.~\ref{eq:SpSmtwist}.
This is the case in Figs.~\ref{fig:CorrelDelta06} and \ref{fig:CorrelDelta-06}.
In Fig.~\ref{fig:CorrelDelta06} the stationary region has a smaller extension, $\sim [-10,10]$. There,
the correlations show a reasonable agreement with Eq.~\ref{eq:SpSmtwist}, but not as precise
as in the two previous cases. The strongest deviations concern the moduli of the correlations: $|\la S^+_{i_0} S^-_{i_0+r}\ra_{\rm NESS}|$ turns out to be larger
than in the ground state in the ferromagnetic case (Fig.~\ref{fig:CorrelDelta06}) while the NESS correlations are larger than in the ground state for antiferromagnetic $\Delta=-0.6$
(Fig.~\ref{fig:CorrelDelta-06}).\footnote{As a caveat we however note that in Fig.~\ref{fig:CorrelDelta-06},
the central region still shows some small magnetization gradient, and is thus not completely stationary.} 
On the other hand, the phase factor  (panels 2 and 4 in Fig.~\ref{fig:CorrelDelta06}) is still a linear function of the the distance $r$.
These deviations reveal some interaction and lattice effects which are not captured by the continuum limit.
Such deviations from Eq.~\ref{eq:SpSmtwist} are  observed when the initial magnetic field approaches the saturation value $h_{\rm sat}$, which also coincides with the regime where
the current does no longer vary linearly with $h$ (Fig.~\ref{fig:conductance}). Equivalently, they occur when the twist period $l=2\pi/\theta$ becomes
less than a few ($\sim 10$) lattice spacings.

\begin{figure}
   \includegraphics[width=8cm]{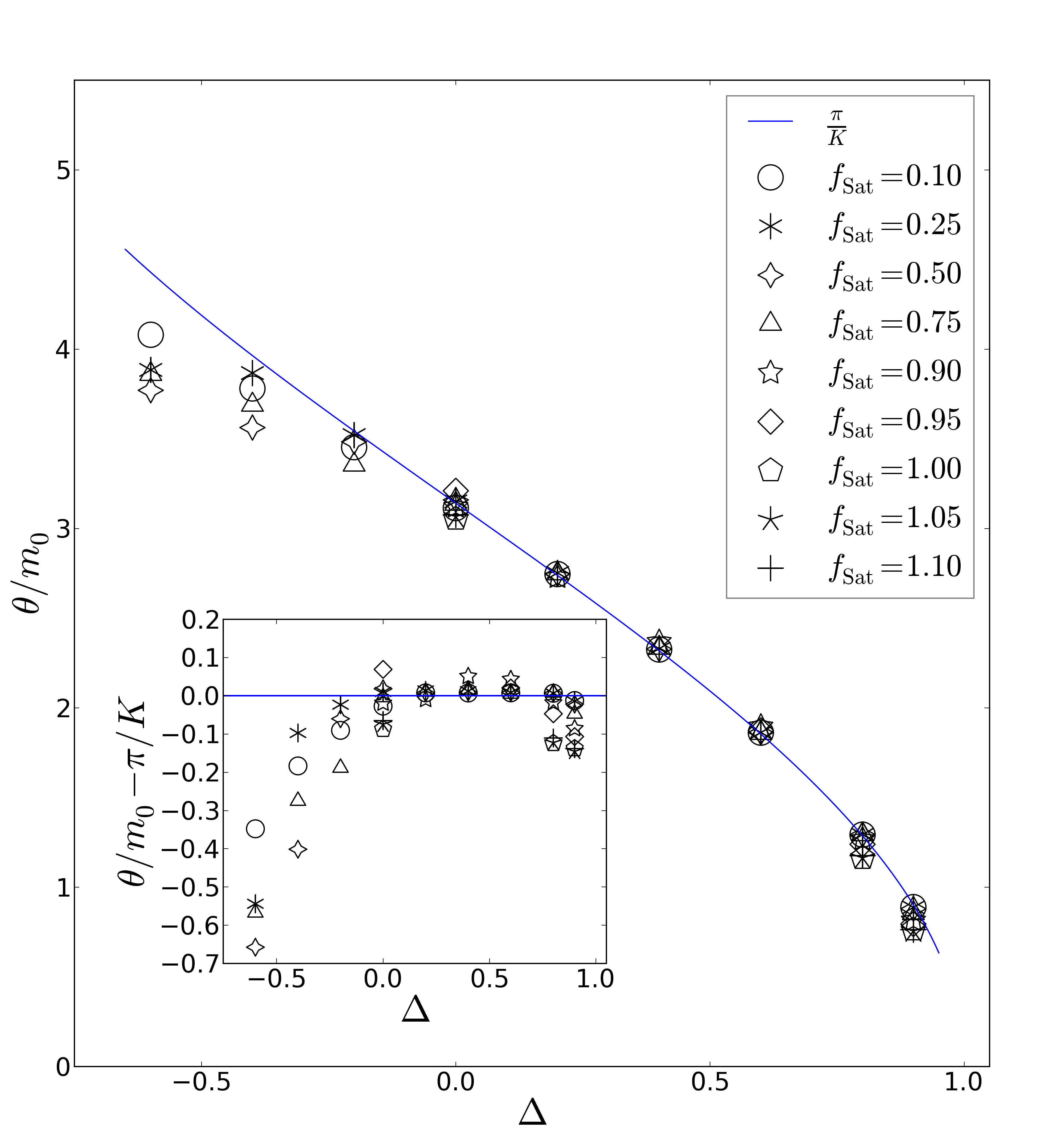}
   \caption[99]{The twist angle $\theta$ describing the steady-state correlation (Eq.~\ref{eq:SpSmtwist}). The blue line corresponds
   to bosonization result of Ref.~\onlinecite{lm10}: $\theta/\pi=\frac{m_0}{K(\Delta)}$ with $K(\Delta)$ given in Eq.~\ref{eq:KDelta}.
   }
\label{fig:LancasterMitraFormula}
\end{figure}

{\it Twist} ---.
As explained above,
the  argument of two-point correlations is linear in $r$ (within the pleateau region) and this allows to define a ``twist'' angle $\theta$.
The results of these fits are displayed in Fig.~\ref{fig:LancasterMitraFormula}, for different values
of $\Delta$ and $h$. The data are plotted as a function of the Luttinger parameter $K$ (Eq.~\ref{eq:KDelta})
and are compared with the bosonization result of Lancaster and Mitra,\cite{lm10} namely:
\begin{equation}
 \theta=\frac{\pi m_0}{K}.
  \label{eq:LM}
\end{equation}
The agreement with our numerics is very good for ferromagnetic (positive) $\Delta$, even for relatively strong currents ($f_{\rm sat}\geq0.9$).
On the other hand, one can see in Fig.~\ref{fig:LancasterMitraFormula} that the agreement somewhat deteriorates for $\Delta\leq0$.
We would however expect Eq.~\ref{eq:LM} to hold for small currents (small $f_{\rm sat}$), whatever $|\Delta|<1$, and it should be exact for 
$\Delta=0$ (whatever $f_{\rm sat}$). We attribute the observed discrepancy to finite-size (and thus finite time) effects, and to the fact that
central region is not exactly steady for  this system size (residual spatial and temporal oscillations or magnetization gradient, etc.).

A possible way to interpret Eq.~\ref{eq:SpSmtwist} is to consider the following unitary transformation:\cite{lsm61}
\begin{equation}
 U(\theta)=\exp\left(i\theta \sum_{r=-L/2}^{L/2-1} r S^z_r\right)
 \label{eq:Udef}
\end{equation}
which satisfies
\begin{equation}
 U(\theta)^\dagger S^+_{i_0} S^-_{i_0+r} U(\theta)= S^+_{i_0} S^-_{i_0+r} \exp\left(ir\theta\right).
\end{equation}
We also have $U(\theta)^\dagger c^\dagger_{i0} c_{i0+r} U(\theta)= c^\dagger_{i0} c_{i0+r} \exp\left(ir\theta\right)$
 since the Jordan-Wigner string commutes with the $S_r^z$ operators and thus commutes with $U$.
Starting from the ground state $|\psi\rangle$  may thus consider the following state:
\begin{equation}
 U(\theta) |\psi\rangle
 \label{eq:UNESS}
\end{equation}
as an approximation to the NESS.
It is the exact NESS for the free fermion point, since $U(\theta)$ boosts all the single particle states
from momentum $p$ to $p+\theta$. It is also possible to check that the results of Ref.~\onlinecite{lm10} (when specialized to the case where
$\Delta$ is not changed during the quench), also imply that Eq.~\ref{eq:UNESS} is the NESS  in the bosonization approximation.
The twist operator can also we written as
\begin{equation}
 U(\theta)=\exp\left(i\theta \sum_{j} \hat r_j \right)
 \label{eq:Uboost}
\end{equation}
where the sum runs over the particles (Jordan-Wigner Fermions) and $\hat r_j$ is the position operator of particle $j$.\footnote{Since $S^z_i=c^\dagger_i c_i-\frac{1}{2}$,
Eq.~\ref{eq:Uboost} and Eq.~\ref{eq:Udef} differ by an irrelevant global phase factor $\exp\left(i\theta \sum_{r=-L/2}^{L/2-1} r/2 \right)=\exp\left(-i\frac{L\theta}{2}\right)$.}
In this form it is clear that $U$ performs a Galilean boost. So, for a system with periodic boundary conditions in the continuum (Galilean invariance),
applying $U$ on an eigenstate gives another eigenstate (with a different energy). If $|\psi\rangle$ is the ground state, $U|\psi\rangle$
sustains some particle current (in the original frame) and is naturally a stationary state (an eigenstate in fact). From this point of view, 
the deviations from Eq.~\ref{eq:SpSmtwist} we observed numerically in the strong current regime are signatures of combined lattice (umklapp) {\it and} interaction effects.

\section{Summary and conclusions}

We have simulated numerically the real-time dynamics of an XXZ spin chain starting at $t=0$ from a state
with different magnetizations on the left and right halves. We have described the shape of the propagating fronts, characterized by two velocities, 
and we have focused on the central region of the chain where an homogeneous current-carrying steady state develops. For moderate value of the magnetization bias,
the value of the current as well as the correlations
in this steady state region turn out to be rather close to the bosonization predictions of Lancaster and Mitra.\cite{lm10}
Indeed, the correlations $\langle S^+_i S^-_{i+r}\rangle$ are close to that of the ground state but multiplied by a phase factor $\exp(i\theta r)$. 
The value of $\theta$ is in good agreement with the continuum limit result (Eq.~\ref{eq:KDelta}).
Also, contrary to most global quenches, the entanglement of a subsystem in the steady state region is not extensive in the subsystem size but it is instead close to that of the ground state
(logarithmic in the length of the subsystem). These properties are easily understood in the free fermion case where the steady state is a ``boosted'' Fermi sea.
However, these are quite remarkable in presence of interactions ($\Delta\ne0$). It is only at large values of the current (or initial bias)
that we begin to observe some deviations from the simple picture of the steady state being a ``boosted'' ground state.
When the current becomes of the order of the maximum current, some corrections to the modulus of the steady state correlations appear (compared to that of the ground state) and
although the twist angle $\theta$ is still well defined, it is no longer given by its bosonisation value (Eq.~\ref{eq:KDelta}).

How to address the combined lattice and interaction effects which are responsible for these deviations from the ``boost'' picture is an interesting problem.
One possible approach could be to include the effects of the non-linear dispersion relation (lattice effects) in a quantum hydrodynamic framework as was done in Ref.~\onlinecite{protopopov}
On the other hand, since the XXZ spin chain is an integrable system, it may be possible to construct the steady state using Bethe Ansatz or integrability techniques, similar to
that of Refs.~\onlinecite{prosen2011,kps13}. It would also be interesting to understand if the steady state can be described in terms of one or several excited eigenstate(s) of
the Hamiltonian.\cite{ce13}
Some results (Loschmidt echo, initial energy distribution, etc.) have already be obtained using the Bethe Ansatz in the case where the $m_0=\frac{1}{2}$ (and $\Delta>1$),\cite{mc10} but
the cases where the reservoirs are partially polarized at $t=0$ is clearly more complex.

\section*{Acknowledgements}
We wish to thank Vincent Pasquier, Benoît Douçot, Masaki Oshikawa, Jean-Marie Stéphan, Jérôme Dubail, Alexandre Lazarescu, Keisuke Tostuka, Edouard Boulat and Aditi Mitra for useful discussions.
The numerical simulations were done on the machine Airain at the Centre de Calcul Recherche et Technologie (CCRT) of the CEA
and on the machine Totoro at the IPhT. This work is supported by a JCJC grant of the Agence National pour la Recherche (project ANR-12-JS04-0010-01). 

\appendix

\section{Steady state in the free-fermion case}
\label{sec:free}

This section is devoted to the XX chain ($\Delta=0$).

\subsection{Long-time limit of correlation functions}

Inspired by some exact results obtained for a specific initial state,\cite{grw10} we look here for the general relation between the NESS and the correlation functions of the {\it initial} state.
We make the assumption that, far on the left and far on the right of the (infinite) chain, the correlations are Gaussian and that the two-point correlations are asymptotically translation invariant.
The left and right parts play the role of reservoirs, and they are completely specified by their occupation numbers
$\langle c^\dagger_p c_p\rangle_{L}$ and $\langle c^\dagger_p c_p\rangle_{R}$. The ground state of Eq.~\ref{eq:tanh}
is of this type but the arguments below apply to more general initial states.\footnote{For simplicity we assume here that the anomalous terms $\langle c^\dagger_p c^\dagger_{-p}\rangle$ vanish.}
This applies, for instance, to two half-chains prepared at different temperatures and/or external magnetic field.

For $\Delta=0$ the Eq.~\ref{eq:HXXZ} reduces (Jordan-Wigner transformation) to a free fermion model, and is diagonalized in momentum space:
\begin{eqnarray}
 H=\sum_p \epsilon_p c_p^\dag c_p\;\;,\epsilon_p=-\cos(p) \\
c_p^\dag=\frac{1}{\sqrt{N}} \sum_n e^{ipn} c^\dag_n
\end{eqnarray}
and the real-space fermion operators obey the following time evolution:
\begin{equation}
 c_m(t)=e^{iHt} c_m(0) e^{-iHt} =\int \frac{dp}{2\pi} c_p e^{i(pm-\epsilon_p t)}.
\end{equation}

We assume that the initial state is defined by a Gaussian density matrix.  It can either be the ground state or the finite-temperature equilibrium density matrix associated to some arbitrary quadratic Hamiltonian.
Then, the free-fermion dynamics will preserve this  Gaussian structure and 
the state (density matrix) will be fully determined by its two-point correlations (Wick theorem). We therefore focus on the correlations:
\begin{eqnarray}
 \langle c^+_m(t) c_n(t)\rangle &=& \int \frac{dp}{2\pi} e^{-i p(m-n)}
  \int \frac{dq}{2\pi}  F_p(q)\nonumber \\
  &&\times e^{it(\epsilon_{p+q/2}-\epsilon_{p-q/2})-i(m+n)q/2} \\
  F_p(q)&=&N \langle c^+_{p+q/2}c_{p-q/2}\rangle_{t=0}
\end{eqnarray}

From now we will {\it assume} that for sufficiently long time the correlations between two sites at finite distance from the origin become independent of $t$.
In other words, we will assume that a steady state region develops. 
The contributions from terms where $q$ is of order one will give rise to a fast oscillating factors since
$\epsilon_{p-q/2}-\epsilon_{p+q/2}$ will be finite. 
These may be important to understand the front shape and dynamics\cite{antal99} but will not contribute to the correlation between sites in the steady state region.
Instead, the correlations which develop in the steady state can only originate from terms where $t(\epsilon_{p-q/2}-\epsilon_{p+v/2})$ is (at most) of order one.
So, following Ref.~\onlinecite{grw10}, we make the following change of variable
\begin{eqnarray}
u&=&2t \sin(p)\sin(q/2) \\
du&=&t \sin(p)\cos(q/2)dq
 \end{eqnarray}
to write 
\begin{eqnarray}
 \langle c^+_m(t) c_n(t)\rangle &=& \int \frac{dp}{2\pi} e^{-i p(m-n)}
  \int \frac{dq}{2\pi}  F_p(q)\nonumber \\
  &&\times e^{2it \sin(p)\sin(q/2)-i(m+n)q/2}  \\
&=& \int \frac{dp}{2\pi} e^{-i p(m-n)}\nonumber \\
  &&\int_{-\infty}^{+\infty} \frac{du}{2\pi t |\sin(p)\cos(q(u)/2)|} \nonumber \\
  &&\times F_p(q(u)) e^{iu-i(m+n)q(u)/2}.
\end{eqnarray}
The steady state assumption now implies that the integral on $u$ is dominated by finite values of $u$ when $t\to\infty$
and, since $m+n$ is finite, we can replace $q(u)$ by zero in the exponential as well as in $\cos(q/2)$. We therefore get some translation-invariant correlations:
\begin{equation}
 \langle c^+_m(t) c_n(t)\rangle _{t\to \infty}=  \int \frac{dp}{2\pi } e^{-i p(m-n)} G(p)
 \end{equation}
 with
 \begin{equation}
  G(p)=\langle c^+_p c_p\rangle_{\rm NESS}= \frac{1}{t |\sin(p)|}\int_{-\infty}^{+\infty} \frac{du}{2\pi}  F_p(q(u))e^{iu}. \label{eq:G0}
\end{equation}
As we will see, although $q(u)\sim u/t$, it would however be incorrect to replace $q(u)$ by zero in $F_p(q(u))$.

For $q=0$, $F_p(0)$ merely measures the occupation number $\langle c^+_p c_p\rangle$ at $t=0$ (conserved quantity) and does not contains the information about the
left/right spatial structure of the initial state, and thus does not have the information about the current that will flow
in the steady state.\footnote{Replacing $q(u)$ by zero in $F_p(q(u))$ would describe the correlations in a different regime, namely that of 
times $t\gg L$ when the fronts have bounced a large number of times at the boundaries of the chain (see Sec.~\ref{ssec:long_times}).}

This information is therefore
encoded in the expansion of $F_p(q)$ in the {\it vicinity} of $q=0$. From the various initial states studied so far (Ref.~\onlinecite{grw10} in particular), it appears
that for small $q$ the initial state correlations $F_p$ are dominated by some {\it pole} with residue noted $f(p)\in \mathcal R$:
\begin{equation}
  F_p(q)\sim  \frac{-i f(p)}{q -i 0^+ {\rm sign}(f(p))}  
  \label{eq:poleFpq}
\end{equation}
This form will be justified in Sec.~\ref{ssec:realspace} and the precise form of residue $f(p)$ will be given (See also Sec.~\ref{a:residue}).

Remarks: i) the ${\rm sign}(f(p))$ insures a positive contribution to $F_p(0)$, as it should since $F_p(0)=N\langle c^+_p c_p\rangle$. ii) On a finite chain the pole divergence at $q=0$ is cut by the system size $N$, so $0^+\sim 1/N$.
iv) In general the behavior close to $q=0$ is the sum of two terms, one with $f^+(p)>0$ and  another one with $f^-(p)<0$.
If $f^+(p)+f^-(p)=0$ we are left with a delta function at $q=0$ (this is realized if the initial state is spatially homogeneous:  $F_p(q)\sim \delta(q)$). 

From the Ansätze above, we get
\begin{eqnarray}
  G(p)&=&{\rm sign}(p) f(p) \nonumber \\
  &&\int_{-\infty}^{+\infty} \frac{du}{2i\pi}  \frac{e^{iu} }{u -i 0^+ {\rm sign}(f(p))\sin(p)},
\end{eqnarray}
which can be computed using a contour in complex plane extending to ${\rm Im}(u)\to +\infty$. Depending on ${\rm sign}(f(p)\sin(p))$ this contour will -- or will not -- enclose
the pole and we get:
\begin{eqnarray}
G(p)&=&{\rm sign}(p) f(p) \theta\left(f(p)p\right) \\
&=& |f(p)| \theta\left(f(p) p\right).
\label{eq:Gp}
\end{eqnarray}
In other words, we find a contribution to the  steady state correlations $\langle c^+_k c_{k'}\rangle_{\rm NESS}$ which is  diagonal in momentum space
and simply related to the pole residue of the initial state correlations.
Remark: for a general dispersion relation, ${\rm sign}(\sin(p))={\rm sign}(p)$ should be replaced by the sign of the group
velocity $\partial_p \epsilon_p$.

\subsection{Reduced density matrix and generalized Gibbs ensemble}
\label{ssec:gge}

The Gaussian nature of the steady state allows to write its density matrix  in terms of the its two-point correlations.
In the present case we find
 \begin{eqnarray}
  \rho_{\rm NESS} &=& \frac{1}{Z}\exp\left(-\sum_p \lambda_p c^\dag_p c_p \right) \\
  Z&=&\prod_p \frac{1}{1+e^{-\lambda_p}}
 \end{eqnarray}
with $\lambda_p=\ln\left(\frac{1-G(p)}{G(p)}\right)$ to insure ${\rm Tr}\left[c^\dag_p c_p \rho\right]=G(p)$.
We note that for the modes $p$ which are completely occupied ($\langle c^\dag_p c_p \rangle=1$) or completely empty ($\langle c^\dag_p c_p \rangle=0$),
$\lambda_p$ is respectively equal to $-\infty$ or $+\infty$.
Although very simple in terms of the fermionic operators, this reduced density matrix corresponds to some rather complex state in terms of the original spin degrees of freedom
(due to the non-local character of the Jordan Wigner transformation). Since the number occupancies $\langle c^\dag_p c_p\rangle$ correspond to all the conserved quantities of the model,
we note that this density matrix is a particular case of the form proposed by Rigol {\it et al.} for generalized equilibrium states of integrable systems.\cite{rigol}

In Ref.~\onlinecite{lm10}, Lancaster and Mitra argued that the generalized Gibbs ensemble (GEE) cannot apply to the present spatially inhomogeneous quenches.
The argument is based on the (correct) observation that
the expectation values of the conserved quantities $I_p$ (here $I_p=c^\dag_p c_p$) are different in the initial state and in the steady state.
The expectation values of $c^\dag_p c_p$ are of course independent of time if the Fourier transform is performed on the whole chain, but the steady state region is only defined on subsystem
of size $l\sim v t$. This  difference between the ``global'' $\langle c^\dag_p c_p \rangle$ and $\langle c^\dag_p c_p \rangle_{\rm NESS}$ is the reason why the steady state occupation numbers $G(p)$
are not those of the initial state $F_p(q=0)$. In our calculation, this difference is the difference between a  $\delta$-function behavior and a pole behavior in $F_p(q)$.
Our point of view is thus that the steady state density matrix does have a GEE form, but the expectation values of the conserved quantities are related in a non-trivial way to that of
the initial state.
So, strictly speaking, the GEE hypothesis does not describe the NESS. The later nevertheless has a density matrix of the GEE form, but there,the  conserved quantities
are not those of the initial state. As discussed in Sec.~\ref{ssec:long_times}, one however recovers the initial state occupation number
$\langle c^\dag_k c_k\rangle$  in the very long time limit after many bounces ($t\gg L$). It is only in this regime that the standard GEE is obtained.

\subsection{Relation to the initial state occupation numbers}
\label{ssec:realspace}

From the argument above we see that a  small fraction of the information contained in the initial correlations $F_p(q)$ ``survives'' in the steady state regime, characterized
by $G(p)$.
In turn, we may ask how to relate this residue information to the initial {\it real space} correlations. As expected on physical grounds, it is the the correlations between sites which are
far on the left or far on the right of the chain which determine the steady state correlations.
We Fourier transform the initial time correlation $F_p(q)$ to real space
\begin{eqnarray}
 F_p(r)&=&\int \frac{dq}{2\pi} F_p(q) e^{-iqr/2} \\
&=&\sum_{\begin{array}{c}n\in \mathbb Z \;{\rm if}\;r\;{\rm even}\\ n\in \mathbb Z+\frac{1}{2}\;{\rm if}\;r\;{\rm odd}\end{array}}
\hspace*{-1cm}\langle c^\dag_{r/2+n}c_{r/2-n} \rangle e^{2ipn}
\end{eqnarray}
to make clear that the relative momentum $q$ is conjugate to the center of mass $r$ of the two sites (Wigner function). 
If we replace $F_p(q)$ by its pole close to $q=0$ we get the behavior of $F_p(r)$ for $r\to \pm \infty$.
Far from the origin ($|r|\gg1$) a pole of the form of Eq.~\ref{eq:poleFpq} corresponds to a step (Heaviside) function in real space:
\begin{equation}
\int_{-\pi}^\pi \frac{dq}{2\pi}\frac{-i f }{q-i0^+ {\rm sign}(f)}e^{-iq r/2} \simeq |f|\theta(- r\;f).
\end{equation}
We conclude that $F_p(q \to0)$ is
is composed by i) a first pole with residue $f_p^+\geq 0$ coming from the  $p$-component of the correlation $\langle c^\dag_{r/2+n}c_{r/2-n} \rangle$
for two sites far on the left ($|r| \to -\infty$),   and ii) a second pole with $f_p^-<0$ coming from the correlations between sites far on the right:
\begin{eqnarray}
f^\pm_p=\pm F_p(r\to \mp \infty).
\label{eq:Ff}
\end{eqnarray}

We consider some initial state which is asymptotically homogeneous in space far on the left ($r\to-\infty$), with $\langle c^\dag_pc_p\rangle=n^-(p)$ and
also asymptotically homogeneous far on the right ($r\to+\infty$), with $\langle c^\dag_pc_p\rangle=n^+(p)$.
In other words: $F_p(r\to\pm\infty)=n^\pm(p)$. Using Eq.~\ref{eq:Gp} and Eq.~\ref{eq:Ff} we obtain
\begin{equation}
  G(p)=F_p(r\to-\infty) \theta (p) + F_p(r\to\infty)\theta (-p).
\label{eq:G}
\end{equation}

We finally have a relation between the steady-state distribution $G(p)$ and the initial time correlations. As expected, the steady state is independent of the details
of the initial state at finite distance from the origin. Only the correlations at $\pm\infty$ matter, which is physically simple to understand since these regions far from the origin
act as reservoirs which drive the central region  out of equilibrium.
\begin{figure}
  \includegraphics[width=9cm]{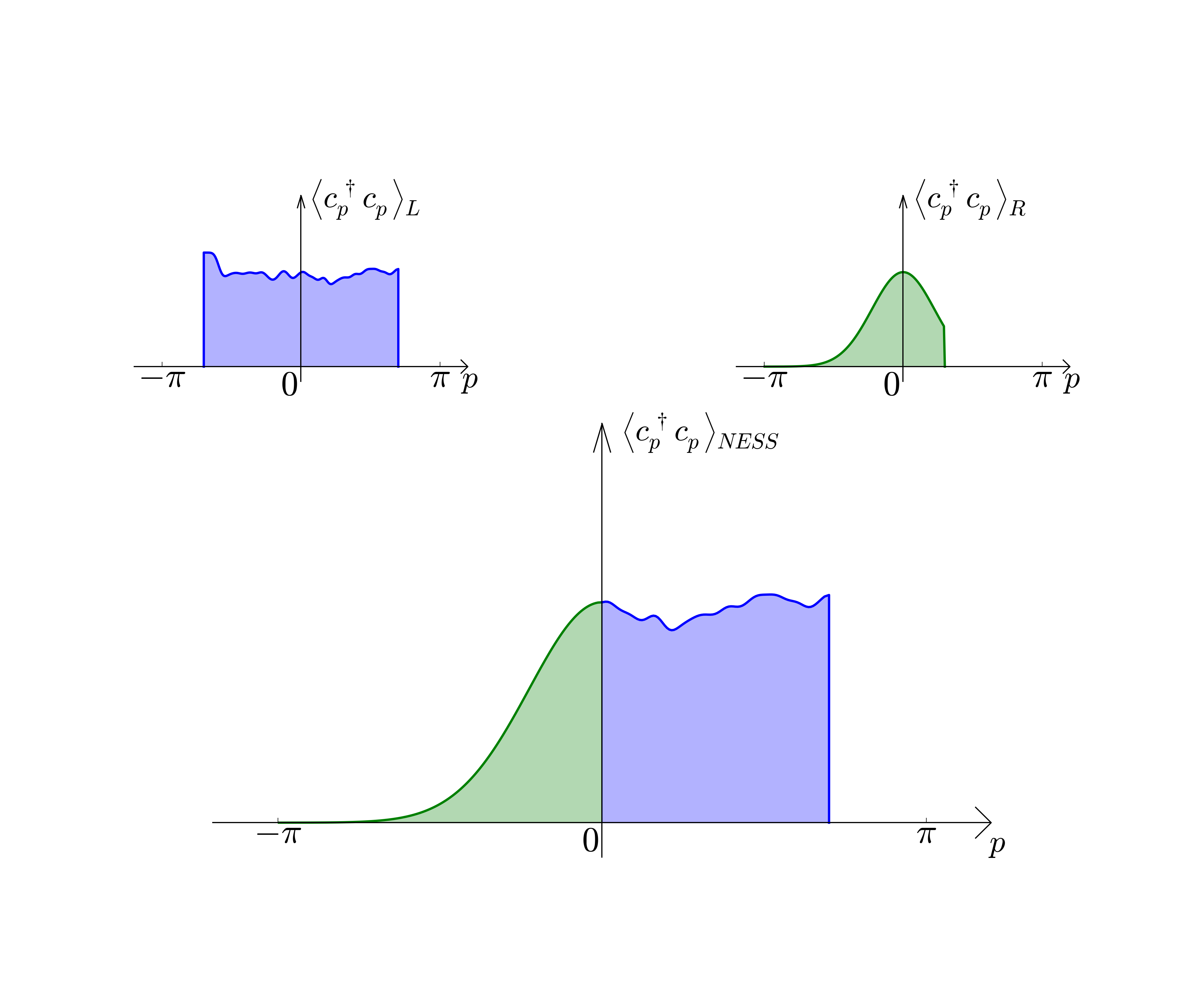}
  \caption[99]{(Color online) Schematic representation of the distribution of the occupations numbers as a function of the momentum $p$ in the region where the non equilibrium steady state is well established $\langle c^\dagger_p c_p\rangle_{NESS}$. The occupation $\langle c^\dagger_p c_p \rangle_L$ on the far left side of the chain, which acts as a \emph{reservoir}, contributes to the distribution of the right movers on the NESS region, whereas $\langle c^\dagger_p c_p \rangle_R$ contributes to the distribution of left movers.}
\label{fig:scheme-occupation}
\end{figure}
If we specialized to the situation where the left (resp. right) half of the chain is prepared at $t=0$ in a
equilibrium state at temperature $T_1$ (resp $T_2$), we find a steady state characterized by a combination of two Fermi distributions (with  temperature $T_1$ for positive momenta, and
with temperature  $T_2$) for negative $p$), as obtained by an explicit -- but somehow more technical -- calculations in Refs.~\onlinecite{ogata02,grw10}.

If we now have two half chains at zero temperature but different chemical potentials (magnetic fields in the spin language). We have two different well defined 
Fermi momenta $k_F^\pm$ far on the left and far on the right . The Eq.~\ref{eq:G} translates into a simple ``rectangular'' distribution:
$G(p)=1$ if $-|k_F^-| \leq p \leq |k_F^+|$ and $G(p)=0$ otherwise. This type of steady state has already been obtained
by an explicit calculation starting from Antal's domain-wall state.\cite{antal99}

We finally note that once $G(p)$ is known, it is a elementary task to compute the particle and energy currents $J$ and $J^E$ flowing in the steady state:
\begin{eqnarray}
 J&=&\frac{1}{2i}\langle c^\dag_{i+1}c_i-c^\dag_ic_{i+1}\rangle \nonumber\\
&=&\int_{-\pi}^\pi \frac{dp}{2\pi} \sin(p) G(p)\\
J^E&=&\frac{1}{4i}\langle c^\dag_{i-1}c_{i+1}-c^\dag_{i+1}c_{i-1}\rangle\nonumber\\
&=&-\int_{-\pi}^\pi \frac{dp}{2\pi} \sin(p)\cos(p) G(p)\\
\end{eqnarray}

\section{Explicit calculation of the pole residue in Antal's initial state}
 \label{a:residue}

In this section we present an explicit calculation of the $q\simeq 0$ behavior of the initial state correlator $F_p(q)$ for a domain wall state.
This initial state is a generalization of that introduced by Antal {\it et al.}\cite{antal99}
We define some annihilation operators for the left (L) and right (R) halves of the chain:
\begin{eqnarray}
 R_k&=&\frac{1}{\sqrt{N}}\sum_{j=1}^N e^{-ikj} c_j \\
L_k&=&\frac{1}{\sqrt{N}}\sum_{j=-N+1}^0 e^{-ikj} c_j \\
 \end{eqnarray}
where $N=L/2$. The initial state $|\phi\rangle$ is then defined as
a tensor product of two states on the left and the right half:
\begin{equation}
 |\phi\rangle=|\phi^L\rangle \otimes |\phi^R\rangle
\end{equation}
with
\begin{equation}
 |\phi^L\rangle = \prod_{k=-k_F^+}^{k_F^+} L^\dagger_k |0\rangle \;\;{\rm and}\;\;
 |\phi^R\rangle = \prod_{k=-k_F^-}^{k_F^-} R^\dagger_k |0\rangle. 
\end{equation}
The two Fermi momenta $k_F^\pm$ define the densities (magnetizations) on the two sides.
Since it is a tensor product, the correlations $\langle\phi| c^\dagger_x c_y |\phi\rangle$ vanish
if $x$ and $y$ are not on the same side. If they are,  we get:
\begin{equation}
\langle\phi| c^\dagger_x c_y |\phi\rangle =
  \left\{
  \begin{array}{cc}
   \frac{1}{N}\int_{-k_F^-}^{k_F^-}\frac{dq}{2\pi} e^{-iq(x-y)}	 & 	x,y\geq 1 \\
   \frac{1}{N}\int_{-k_F^+}^{k_F^+}\frac{dq}{2\pi} e^{-iq(x-y)}	& 	x,y<0
  \end{array}
  \right.
\end{equation}
Now we Fourier transform these correlations to get $F(k,k')=\langle \phi | c^\dagger_k c_{k'} | \phi \rangle$:
\begin{eqnarray}
 F(k,k')&=&F^L(k,k')+F^R(k,k')\\
  F^R(k,k')&=&\sum_{x,y\geq1}\int_{-k_F^-}^{k_F^-} \frac{dq}{2\pi} e^{-iq(x-y)+ikx-ik'y} \\ 
\end{eqnarray}
and $F^L(k,k')$ is simply obtained by changing the sum into $\sum_{x,y\leq0}$.
In the following we restrict the discussion to $F^R$ for brevity.
The  sum over $x\geq 0$ is made convergent in the thermodynamic limit  by changing
$k$ into $k+ i0^+$. In the same way, we regularize the sum over $y$ by $k'\to k' +i0^-$.
These sums and the integration over $dq$ can then be done and lead to
\begin{eqnarray}
 F^R(k,k')&=&\frac{1}{2i\pi}\frac{1}{1-e^{-i(k-k')}} \left[  \begin{array}{c}\\ \\ \\\end{array}\right.
\\
\ln\left(
  \frac{e^{ik_F^-}-e^{-ik}}{e^{-ik_F^-}-e^{-ik}}
\right) &-&
\ln\left(
  \frac{e^{ik_F^-}-e^{-ik'}}{e^{-ik_F^-}-e^{-ik'}}
\right)
\\
&-&i\pi\;{\rm if}\;k\in[-k_F^-,k_F^-]\\
&-&i\pi\;{\rm if}\;k'\in[-k_F^-,k_F^-]
\left.\begin{array}{c}\\ \\ \\\end{array} \right]
 \end{eqnarray}
Now we want to extract the pole contributions when $k$ is close to $k'$. The two $\ln$ vanish in this limit
and we are left with
\begin{eqnarray}
 F^R(k,k') \simeq \left\{
 \begin{array}{cc}
  \frac{i}{k-k'+i0^+} & k\in[-k_F^-,k_F^-] \\
  0 & k\notin[-k_F^-,k_F^-]
 \end{array}
  \right. .
\end{eqnarray}
The expression above corresponds to Eq.~\ref{eq:poleFpq}
with $f(p)=f^R(p)$:
\begin{equation}
 f^R(p)=\left\{
 \begin{array}{cc}
  -1 & p\in[-k_F^-,k_F^-] \\
    0 & {\rm otherwise}
 \end{array}
  \right.
  \label{eq:fR}
\end{equation}
In a similar way, we get another pole coming from $F^L(k,k')$:
\begin{equation}
 f^L(p)=\left\{
 \begin{array}{cc}
   1 & p\in[-k_F^+,k_F^+] \\
    0 & {\rm otherwise}
 \end{array}
  \right.
  \label{eq:fL}
\end{equation}
Combining Eq.~\ref{eq:Gp} with Eqs.~\ref{eq:fR}-\ref{eq:fL}, we finally obtain the steady state occupation numbers:
\begin{equation}
G(p)= \left\{
\begin{array}{cc}
 1 & k\in[-k_F^-,k_F^+] \\
 0 & {\rm otherwise}
\end{array}\right.
\label{eq:GpFinal}
\end{equation}



\end{document}